\documentclass[%
 reprint,
 amsmath,amssymb,
 aps
]{revtex4-2}

\usepackage{graphicx}
\usepackage{dcolumn}
\usepackage{bm}
\usepackage{hyperref}
\usepackage{booktabs}
\usepackage{aas_macros}
\usepackage{xcolor}

\begin{document}


\title{Measurability of precession and eccentricity for heavy binary-black-hole mergers}

\author{Yumeng Xu}
 \affiliation{%
 Physik-Institut, Universit\"{a}t Z\"{u}rich, Winterthurerstrasse 190, 8057 Z\"{u}rich, Switzerland
}%
\author{Eleanor Hamilton}%
\affiliation{%
 Physik-Institut, Universit\"{a}t Z\"{u}rich, Winterthurerstrasse 190, 8057 Z\"{u}rich, Switzerland
}%

\date{\today}

\begin{abstract}
Gravitational wave detections offer insights into the astrophysical populations of black holes in the universe and their formation processes. Detections of binaries consisting of black holes lying outside the bulk distribution of the astrophysical population are particularly intriguing. 
In this study, we perform an injection analysis within the intermediate-mass black hole range, utilizing the NR surrogate model NRSur7dq4 and a selection of NR waveforms from the SXS and RIT catalogues. Our investigation focuses on the detectability of precession and its potential degeneracy with eccentricity, especially for short signals with only a few cycles in band. 
While total mass, mass ratio, and $\chi_\mathrm{eff}$ are generally well recovered, the recovery of $\chi_\mathrm{p}$ is largely limited, and noise significantly impacts the recovery of some parameters for short signals.
We also find that eccentricity lower than 0.2 is insufficient to mimic precession in parameter estimation when assuming a quasi-circular signal. Our results suggest that a certain degree of precession is necessary to produce evidence of high precession in parameter estimation, but it remains challenging to conclusively determine which effect is responsible for the high precession observed in events like GW190521. We emphasize the importance of caution when interpreting properties of a binary from short signals and highlight the potential benefits of future third generation detectors and eccentric waveform models for more exhaustive exploration of parameter space.
\end{abstract}

\maketitle

\section{Introduction}
Over the past few years, the LIGO-Virgo-KAGRA (LVK) network of gravitational wave detectors has successfully completed three runs, resulting in the detection of over 90 gravitational wave events \cite{LIGOScientific:2021djp}. With the upcoming fourth run (O4), the network is expected to detect 3 times more events \cite{LIGOScientific:2021psn}.
These detections bring us to a new era of astrophysics and provide us with a new possibility to answer questions about the astrophysical makeup of the universe.

There are two particular questions of interest: the properties and population of intermediate-mass black holes (IMBHs) and the existence of black holes within the pair-instability supernova (PISN) mass gap.
IMBHs have masses in the range $10^2 - 10^5 M_\odot$, bridging the gap between stellar BHs and supermassive BHs. The understanding of the properties and population of IMBHs might provide the missing link to explain the formation of supermassive black holes \cite{IMBH}.
The PISN mass gap refers to the range of black hole masses between $65 M_\odot$ and $120 M_\odot$, which is disfavoured to be formed from stellar origin. Within this gap, when a star with a helium core reaches the end of central carbon burning and collapses inwards, there will be a complete disruption of the core and no compact object will remain\cite{PISN1, PISN2, PISN3, PISN4, PISN5}. 
Alternative formation channels for BHs within this gap --- including second-generation BHs, stellar mergers in young star clusters, black hole mergers in active galactic nucleus disks, primordial black holes, and gas accretion --- have been widely discussed \cite{2020ApJ...900L..13A, Safarzadeh:2020vbv, PhysRevLett.126.051101, Holgado:2020imj, Palmese:2020xmk, Sedda:2021abh, Rice:2020gyx}.

Distinguishing the formation channels relies on accurate measurements of key properties of the binary, in particular precession and eccentricity\cite{Rodriguez:2016vmx, Steinle:2020xej}. A quasi-circular binary black hole (BBH) is characterised by its intrinsic parameters; the total mass $M_\mathrm{tot}$ of the binary system and mass ratio $q=m_2/m_1$, where $m_1>m_2$ is the mass of the primary black hole, and the spins $\vec{S}_1$ and $\vec{S}_2$.
If we consider a binary on an eccentric orbit then we add at least one additional intrinsic parameter; the eccentricity $e$.

Spin-induced precession occurs when the spins are not aligned with the orbital angular momentum. Precession introduces modulations of the GW amplitude and phase \cite{Lang2006, Schmidt:2012rh}. From post-Newtonian theory, we can construct an effective spin $\chi_\mathrm{eff} = (\vec S_1/m_1 + {\vec S_2}/{m_2} ) \cdot \hat L / M_\mathrm{tot}$~\cite{Ajith:2009bn, Santamaria:2010yb} which characterizes the impact of the components of the spins aligned with the orbital angular momentum of the binary $\hat L$. The effective spin can be measured better than the individual spins \cite{2011PhRvL.106x1101A, 2017PhRvD..95f4053V}. 
The components of the spins which lie perpendicular to $\hat L$ cause $\hat L$ to precess about the total angular momentum of the binary, $\hat J = \hat L + \hat S_1 + \hat S_2$. 
The dominant effects of precession on the GW signal can be characterized by the effective precession spin
$\chi_p = \max(A_1 S_{1\perp}, A_2 S_{2\perp}) / (A_1 m_1^2)$
where $A_1 = 2 + 3q / 2$ and $A_2 = 2 + 3 /(2q)$~\cite{Schmidt:2014iyl}. 
$S_{1\perp}$ and $S_{2\perp}$ are the magnitudes of the components of the spins that lie in the orbital plane.

We can use GW detections to build up a picture of the population of BHs in the Universe.
The LVK network has detected a number of events which fall into either the population of IMBHs or the PISN mass gap \cite{LIGOScientific:2021usb}. Most of the detection are marginal cases. Among these detections, GW190521 \cite{Abbott2020,2020ApJ...900L..13A} is particularly interesting as it is the first-ever detection of a black hole in PISN mass gap merging with another black hole to form an IMBH and it is also the only one that's not marginal.
There have only been three signals reported by the LVK so far which show any support for precession: GW190412, GW190521 and GW200129\_065458 \cite{LIGOScientific:2020stg, Abbott2020, LIGOScientific:2021djp, Hannam:2021pit, Payne:2022spz}. Eccentricity can also cause modulations in the amplitude and phase of the signal \cite{Huerta:2017kez} which we haven't had highly confident detection.

Difficulties in accurately determining the properties of an IMBH system arise because this kind of heavy BBH event has a very short duration and only has a few cycles in the frequency range to which aLIGO and aVirgo are most sensitive \cite{LIGOScientific:2014pky, VIRGO:2014yos}. 
For instance, GW190521 only has approximately 0.1s in duration and around 4 cycles in the frequency band 30-80Hz \cite{Abbott2020}.
The early inspiral and even part of the late inspiral were outside of the sensitivity band of the detectors at the time at which the signal was detected. 
This kind of short signal leaves opens the possibility that the inferred properties of the binary may not be accurate, since such a short signal will have features that can be produced by a number of degenerate parameters.
Consequently, the interpretation of the source as a quasi-circular compact binary coalescence consisting of inspiral, merger and ringdown phases is not certain.
In the case of GW90521, it has been seen that the signal can be shown to be consistent with that originating from a number of less likely sources~\cite{Nitz:2020mga, Romero-Shaw:2020thy, Gayathri2022, Bustillo:2020syj, Bustillo:2020ukp}. 
The parameter estimation (PE) is also likely to be disproportionately affected by noise in the detector. Thus, checking the measurability of precession and eccentricity is crucial to be confident of any astrophysical inferences drawn from gravitational wave events within the IMBH mass range.

In this paper, we will consider only signals which have come from a compact binary coalescence (CBC) source, though not necessarily a quasi-circular one. We will examine the measurability of precession by analyzing injected waveforms with known properties similar to those of GW190521. The paper is organized as follows.
In Sec.~\ref{sec:Methodology}, we give an introduction to the methodology we will use in this paper. 
In Sec.~\ref{sec:results}, we investigate the inferences that can be made about signals from quasi-circular binaries. In particular, we consider the robustness of these inferences with respect to noise in the detector. 
These investigations are extended to include eccentric systems in Sec.~\ref{sec:ecc}. However, since we don't have an eccentric waveform model for the merger, we will only check the bias of precession from the eccentric injection.
The detectability from future detectors is also discussed in Sec.~\ref{sec:res:future_det}. 
Finally, we give our conclusions and discussions from these analyses in Sec.~\ref{sec:discussion}.

\section{Methodology}\label{sec:Methodology}

To study these kinds of events by injection, we first must choose the most appropriate waveform model for our investigation. The motivation for our choices is outlined in Sec.~\ref{sec:metho:wf}. The methods employed to generate the waveforms and perform the injections are described in Sec.~\ref{sec:Methodology:injection}. We discuss the setup employed for the PE in Sec.~\ref{sec:Methodology:pe}.
The techniques used to further analyse the results of the PE are then outlined in Secs.~\ref{sec:Methodology:js} and~\ref{sec:Methodology:psnr}.

\subsection{Waveform models}\label{sec:metho:wf}

A number of precessing waveform models have been developed to date \cite{IMRPhanomPv2, IMRPhenomXPHM, IMRPhenomPv3HM, NRSur7dq4, SEOBNRv3, SEOBNRv4PHM1, PhanomPNR}.
In our analysis, we use one of these models, NRSur7dq4~\cite{NRSur7dq4}.
NRSur7dq4 is a merger-ringdown model based on a catalogue of 1528 NR simulations taken from the SXS Collaboration catalog.
These simulations, and thus the model, cover the 7-dimensional parameter space of quasi-circular binaries with generic spins.
However, since it is based solely on numerical relativity (NR) waveforms, it is limited in application to signals containing minimal inspiral-- specifically, high mass systems.
In our studying cases, 
which only contains the merger-ringdown stage, NRSur7dq4 is the most accurate waveform available and therefore the most appropriate for use with in-depth studies of a given event, such as presented in this paper \cite{IMRPhenomXPHM,PhanomPNR}.

When considering quasi-circular binaries, we use this model both for generating our simulated signal and for the PE we perform on this signal. 
Using the same waveform model for both parts of the analysis reduces the possibility that any effects we see are due to waveform systematics.

When considering eccentric signals, there are several eccentric waveform models available \cite{Moore:2016qxz, Tanay:2016zog, Huerta:2017kez, Chiaramello:2020ehz,Ramos-Buades:2021adz} 
which one could use to produce the injected signal. However, these waveform models are insufficient for our study
since the majority of our signal is composed of the merger and ringdown while these models are most accurate in the inspiral regime.
Instead we use waveforms from the SXS Collaboration catalog of binary black hole simulations \cite{Boyle:2019kee} and the RIT Catalog of Numerical Simulations \cite{Healy2022wdn} to simulate the signal from an eccentric binary. 
We recover these injections using NRSur7dq4 as in the quasi-circular case, since to date all PE performed by the LVK has used quasi-circular models.

\subsection{Generation of injected signals}\label{sec:Methodology:injection}

We simulate a GW signal with known source parameters using either NRSur7dq4 or an appropriate NR waveform.
This signal is then used to create a GW ``injection'', consisting of the GW strain as seen by the detectors.
Inspired by the only known precessing IMBH detection, we choose the parameters which form the basis of our injections
the max-loglikelihood parameters from the posterior by the LVK for GW190521  \cite{posterior_samples} and listed in Tab.~\ref{tb:maxL}.
These are the values for the waveform that were found to agree best with the event data in the original LVK analysis.
We vary the selection of these parameters individually.
In what follows, we will refer to the injection in zero noise using exactly these parameters as the ``maxL injection''.

\begin{table}[t]
\caption{\label{tb:maxL} maxL values of GW190521 from the data release \cite{posterior_samples} associated with Ref.~\cite{Abbott2020}.
}
\begin{ruledtabular}
\begin{tabular}{@{}lll}
Symbol & Name & maxL value \\
\colrule
$m_1 (M_\odot)$ & mass of primary black hole      & 147.767     \\
 & (detector frame) & \\
$m_2 (M_\odot)$ & mass of secondary black hole     & 121.063  \\
 & (detector frame) & \\

$\vec{S}_1$ & spin of primary black hole                      & [0.008, -0.071, 0.054]    \\
$\vec{S}_2$ & spin of secondary black hole  & [0.808, -0.358, 0.143]   \\
$\theta_\mathrm{jn} (\mathrm{rad})$  &  angle between line of sight      & 0.953       \\
 & and total angular momentum  & \\
$\iota (\mathrm{rad})$      &  inclination        & 0.834       \\
$\psi (\mathrm{rad})$       &  polarisation      & 2.382       \\
$\alpha (\mathrm{rad})$  & right ascension               & 0.164       \\
$\delta (\mathrm{rad})$ & declination                  & -1.143      \\
SNR      & signal to noise ratio                     & 15.403      \\
$\phi (\mathrm{rad})$       & coalescence phase     & 0.004       \\
$\chi_\mathrm{eff}$   &   effective spin    & 0.094       \\
$\chi_\mathrm{p}$     &   effective precessing spin     & 0.704       \\
$d_L (\mathrm{parsecs})$      & luminosity distance    & 2941.027  \\
\end{tabular}
\end{ruledtabular}
\end{table}

The GW signal $h$ from a CBC can be written as a combination of two polarisations $h_+$ and $h_\times$ as $h=h_+ - i h_\times$. It can be decomposed into modes $h_{\ell m}$ via~\cite{Thorne:1980ru}
\begin{equation}
    h\left(\iota,\phi_0\right)
    = \sum_{\ell=2}^{\infty} \sum_{m=-\ell}^{\ell} 
    h_{\ell m}
    \,^{-2}Y_{\ell m}\left(\iota,\phi_0\right)
\end{equation}
where the $^{-2}Y_{\ell m}$ are a basis of spin-weighted spherical harmonics and $\left(\iota,\phi_0\right)$ give the direction of the radiation in the source frame. The signal itself can be considered as a function of time or frequency and depends on the parameters of the binary.

Both the NRSur7dq4 model and the NR waveforms supply the GW modes $h_{\ell m}$, which are recombined to produce the GW strain $h$ using \texttt{get\_td\_waveform} from \texttt{PyCBC} \cite{pycbc2} and  \texttt{SimInspiralChooseTDWaveform} from \texttt{LALSimulation} \cite{lalsuite} as part of the NR Injection Infrastructure \cite{Schmidt:2017btt} respectively.
The $h_+$ and $h_\times$ polarisations are then projected onto a detector network using the appropriate detector response functions $F_+(\alpha, \delta, \psi)$ and $F_\times(\alpha, \delta, \psi)$ using \texttt{Detector.project\_wave} from \texttt{PyCBC} \cite{pycbc2}. 
$\alpha, \delta, \psi$ are as defined as in Tab.~\ref{tb:maxL}.
This, combined with a given noise realisation, gives us the signal as would be seen by the detectors. This signal forms our ``injection'', upon which we perform our analysis.

In what follows, if the choice of noise realisation is left unspecified the injections will be zero-noise. The purpose of these injections is to see an unbiased PE recovery, since a particular realisation of Gaussian noise might cause a bias in the PE, especially for the kind of short signal under investigation.
A zero noise injection can be treated as the average of all Gaussian noise realisations.

For injections where we wish to investigate the effect of noise on a signal of this kind, we generate coloured Gaussian noise from the power spectral density (PSD) for a given detector. 
The addition of Gaussian noise to the signal is intended to allow for an assessment of the robustness of the PE performed on these signals.
We anticipate that this will give a conservative limit on the effect of noise on the signal since much stronger effects are possible from real detector noise.

\subsection{Bayesian Parameter Estimation}\label{sec:Methodology:pe}

Having produced a simulated GW signal of known source parameters, we perform the same analysis on our simulated signal as is performed on detected signals by the LVK. 
This analysis is intended to determine the parameters of the source.

Given the data, the posterior distribution for a given black hole parameter is given by 
\cite{maggiore2007gravitational}
$$P(\mathrm{parameter}| \mathrm{data}) \propto P(\mathrm{data}| \mathrm{parameter}) \times P(\mathrm{parameter})\,,$$
where the first term on the right-hand side is the likelihood, which is the probability of the data given the parameter. The second term is the distribution we assume for a given parameter, known as the prior. The prior should be astrophysically motivated.
Our assumptions for the priors employed in this study are identical to those used in the LVK analysis of GW190521 \cite{2020ApJ...900L..13A}, except the priors are uniform on chirp mass (70-150 $M_\mathrm{tot}$) and mass ratio (0.17-1.0) to cover the parameter space of IMBH event.

We used the PE code \texttt{Parallel-Bilby} \cite{bilby, Romero-Shaw:2020owr, pbilby} with the nested sampler \texttt{Dynesty} \cite{dynesty} in order to analyse each of our injected signals.

\begin{figure}[t]
    \includegraphics[width=\linewidth]{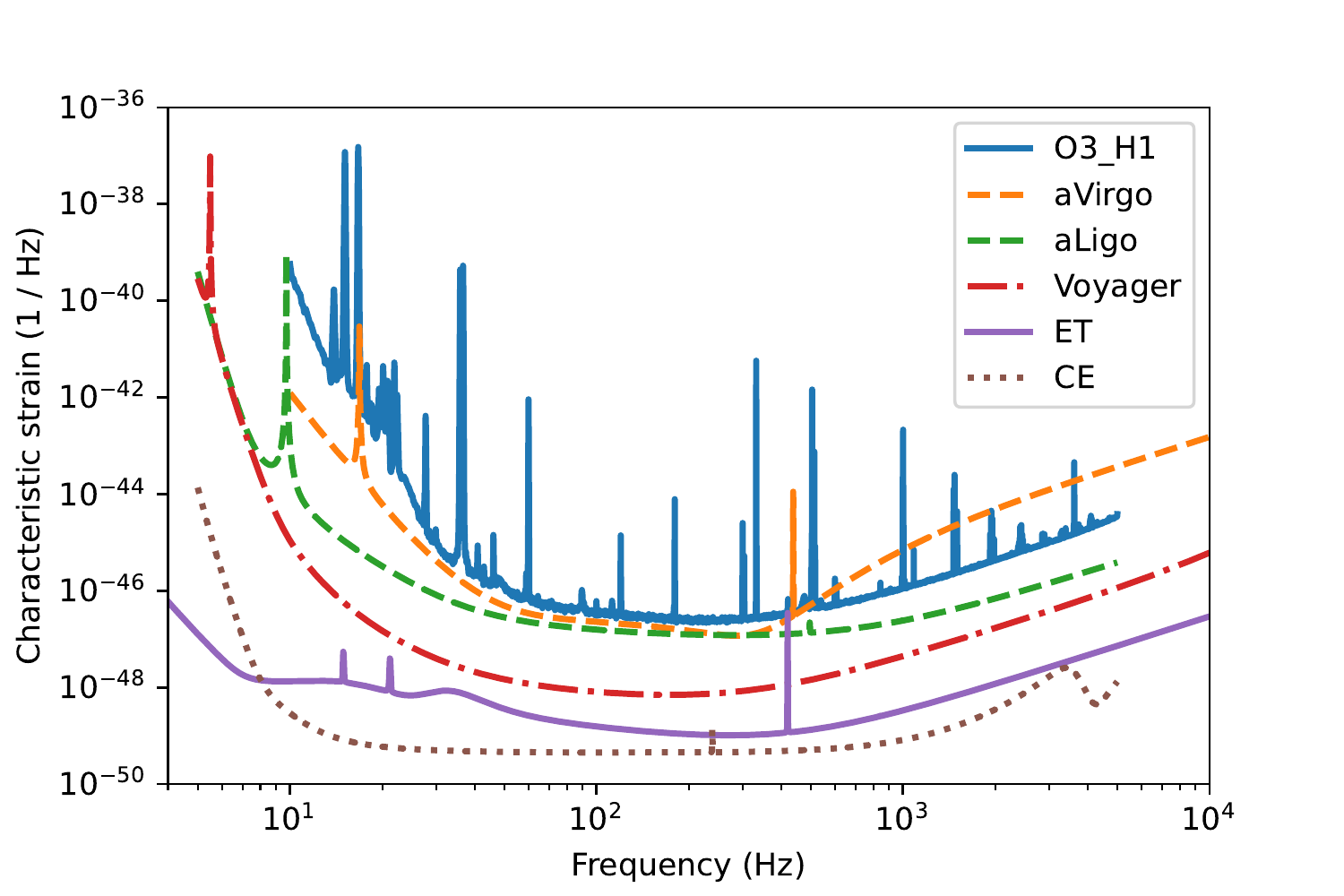}
    \caption{\label{fig:sen_curv}Sensitivity curves of different interferometers. O3\_H1 is the PSD of LIGO Hanford when GW190521 was detected. aLigo and aVirgo curves are the projected curves at design sensitivity. The other curves are proposed sensitivity curves for future detectors.}
\end{figure}

The PSDs of the characteristic sensitivity curves are used to simulate the stationary Gaussian noise and compute the likelihood function.
The various characteristic sensitivity curves we used in our analysis are shown in Fig.~\ref{fig:sen_curv}. 
The PSDs used in the bulk of the analysis come from the LVK results for GW190512 (referred to as event PSDs) ~\cite{posterior_samples} which was calculated from the data by the \texttt{BayesLine} algorithm~\cite{Littenberg:2014oda}. We choose this PSD to make the analysis closer to realistic. 
The curve for the detector H1 is shown, labelled \emph{O3\_H1}.
We also consider sensitivity curves for aLIGO and aVirgo~\cite{ligoasd,virgoasd} (which will be referred to as design PDSs) as well as future detectors Voyager, Cosmic Explorer (CE) and Einstein Telescope (ET)~\cite{sensitivity_curve} in the discussion surrounding future detections in Sec.~\ref{sec:res:future_det}.

\subsection{Jensen–Shannon divergence}\label{sec:Methodology:js}

Having performed PE on each of our injected signals, we wish to be able to perform a quantitative comparison of the results of our various analyses.
We quantify the deviation between two distributions using the Jensen-Shannon (JS) divergence~\cite{1207388}.
The JS-divergence is a variant of the Kullback-Leibler divergence $D_\mathrm{KL}$ \cite{10.1214/aoms/1177729694}. For discrete probability distributions P and Q on the same probability space $\chi$
\begin{equation}
    D_\mathrm{JS} (P || Q) = \frac{1}{2} D_\mathrm{KL} (P || M) + \frac{1}{2} D_\mathrm{KL} (Q || M),
\end{equation}
where $M = (P + Q) / 2$ and
\begin{equation}
	D_\mathrm{KL} (P||Q) = \sum_{x \in \chi} P(x) \log\left(\frac{P(x)}{Q(x)}\right).
\end{equation}

$D_\mathrm{JS} (P || Q)$ is a normalized quantity. The maximum value is dependent on the base; under base 2, which we employ here, the Jensen–Shannon divergence is bounded by 1. 
The lower the JS divergence, the closer the two distributions.
When the two distributions are identical, the divergence is $D_\mathrm{JS} (P||P) = 0$. A value of $D_\mathrm{JS} < 0.05$ is generally taken to mean that the two distributions are in good agreement \cite{LIGOScientific:2018mvr}.

\subsection{Precessing SNR}\label{sec:Methodology:psnr}

One of the most interesting questions about IMBH binaries and those containing BHs in the PISN mass gap is the degree to which it is possible to determine whether the binary is precessing from the detected signal.
In this analysis, we therefore require a statistical tool to assist us in identifying whether we can claim a detected GW to be unambiguously precessing. To do so, we use the precessing SNR~\cite{Fairhurst2020, Green2020}, which is used to calculate whether the precession in a binary can be confidently inferred from detection with a given PSD. The precessing SNR is based on the two-harmonic approximation~\cite{Fairhurst2020}. In this decomposition, the precessing waveform strain is decomposed into five harmonics. Each harmonic is a non-precessing signal. Therefore, if we want to detect the precessing signal, there should be at least two dominant harmonics $h^0$ and $h^1$. 
Thus the precessing SNR $\rho_\textrm{p}$ is defined by the minimal SNR of  $h^0$ and $h^1$
\begin{equation}
    \rho_\textrm{p} = \min(|\mathcal{A}_0 h^0|, |\mathcal{A}_1 h^1|),
\end{equation}
where $\mathcal{A}_0$ and $\mathcal{A}_1$ are the amplitude of each harmonic.

A value of $\rho_\textrm{p}>2.1$ indicates the presence of the second harmonic in the signal at a sufficient level that it is not likely to have arisen from noise fluctuations \cite{Fairhurst2020, Green2020, Hoy:2021dqg}.
$\rho_\textrm{p}=2.1$ was therefore proposed to be a threshold above which a signal may be considered to be unambiguously precessing.
However, we note that in Ref.~\cite{Pratten:2020igi} it was found that for a comparable mass ratio event, the threshold value of $\rho_\mathrm{p}$ may have to be higher than 2.1 in order to confidently claim a detection of precession.
We therefore consider 2.1 to be a conservative estimate of the threshold below which a detection cannot be considered to be unambiguously precessing.

\section{Precession Detectability in IMBH Merger}\label{sec:results}

To gain a deeper insight into the robustness of detecting strong precession in IMBH binaries, we performed a wide-ranging injection study. We first validated the existing understanding of the limitations in detecting precession for low mass ratios and events with low to moderate signal-to-noise ratios. We then looked at the impact of detector noise on such a short signal. Finally, we considered the conditions under which one might be able to detect precession in an IMBH event.

\subsection{Exploring GW190521-like Injections and Varying Total Mass} 

First, we consider a system inspired by GW190521, the only confidently-detected IMBH system to date. The properties of this system are listed in Tab.~\ref{tb:maxL}.
We injected this signal as described in Sec.~\ref{sec:Methodology:injection}.

The recovered posteriors for $M_\mathrm{tot}$, $q$ and $\chi_\mathrm{eff}$ are centred on the injected values, which suggests accurate recovery of these parameters in zero noise. 
$\chi_\mathrm{eff}$ is particularly well-recovered, with a narrow posterior.

\begin{figure}[t]
    \includegraphics[width=\linewidth]{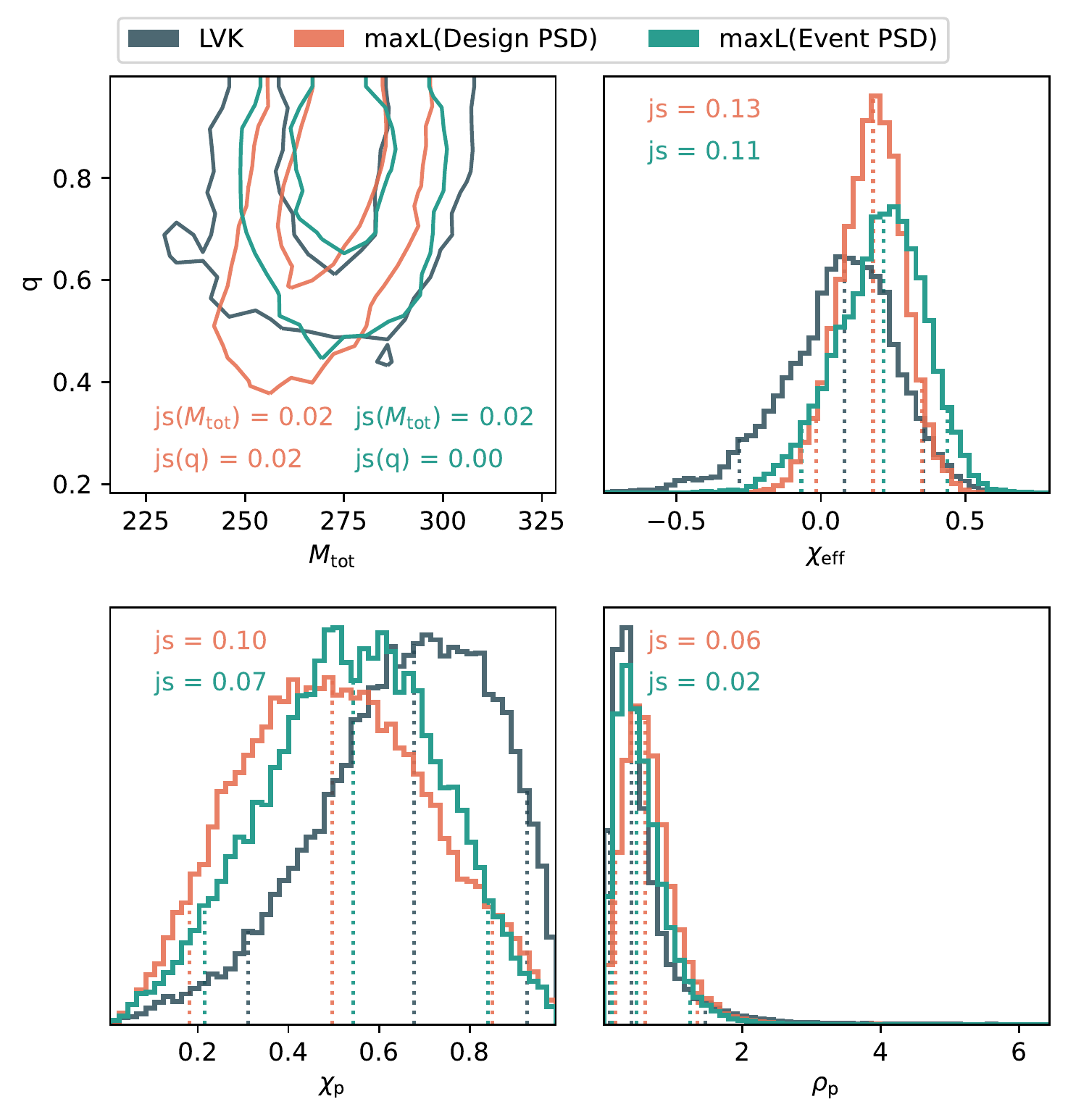}
    \caption{\label{fig:maxL} Results of maxL injection. The blue lines represent the posterior samples from the LVK event data set. The orange lines are from the maxL injection. The vertical dashed lines show the median (50\%) and the bounds of the 90\% credible interval and the contours show the 50\%, and 90\% credible interval. The contours of total mass $M_\mathrm{tot}$ and mass ratio $q$ show a good match. However, the effective spin and effective precession spin show significant deviation from the LVK results.}
\end{figure}

However, the posterior for $\chi_\textrm{p}$ is not centred on the injected value, although this value is included at 90\% confidence.
In fact, the JS divergence for $\chi_\textrm{p}$ with respect to the prior is 0.03 bits and 0.06 bits when considering the injection using the design PSDs and event PSDs respectively. We can therefore see that for the injection we are simply recovering the prior for $\chi_{\textrm{p}}$ and we do not get any information about the amount of precession in the signal. From this, we can conclude that even when we have a highly precessing binary like the injected signal, we do not expect to be able to measure the precession in the signal for such a low mass ratio, low SNR system. 

This conclusion is corroborated by the calculation of the precession SNR $\rho_{\textrm{p}}$ for the injected signal. At 95\% confidence, the precession SNR of the injection is below 2.1, the minimum threshold required to claim an unambiguous detection of precession. The low value of the precession SNR for the event implies we would not expect to be able to make a confident detection of precession here. 

When comparing the maxL injections with the results for GW190521, we find that the JS divergence between the posteriors of the LVK results and the maxL injections is less than 0.02 bits for both $M_\mathrm{tot}$ and $q$. This similarity indicates a good match between the injection and the values measured for the event, meaning we have likely injected a signal similar in total mass and mass ratio to our example IMBH binary.

However, our spin posteriors are not so similar. 
For example, we do not see the support for the negative $\chi_\mathrm{eff}$ values seen in the event posterior. This broadening of the posterior is likely due to detector noise. The high value of the peak in the $\chi_\textrm{p}$ posterior distribution and significant deviation from the prior (a JS divergence of 0.16 bits) seen for GW190521 is not seen for our injection.
It is therefore difficult to confidently claim our injected signal has the same in-plane spin as our example binary.
Further, since we do not reproduce posteriors indicating a high degree of precessing in the signal, the indications of this seen for GW190521 are unlikely to be due to an equivalent or smaller degree of precession than in our injected signal and more likely to be due to other factors. 

It is unlikely that the fiducial binary is exactly the same as the one that produced GW190521, even though it was generated using the "best matching signal." Therefore, we need to explore other parameters to see what the PE from similar IMBH binaries looks like. Since the results obtained using the event PSD and those obtained with the design PSD show negligible differences, we will only use design PSD results as the maxL injection for comparison in the following sections.

\begin{figure}[t]
    \includegraphics[width=\linewidth]{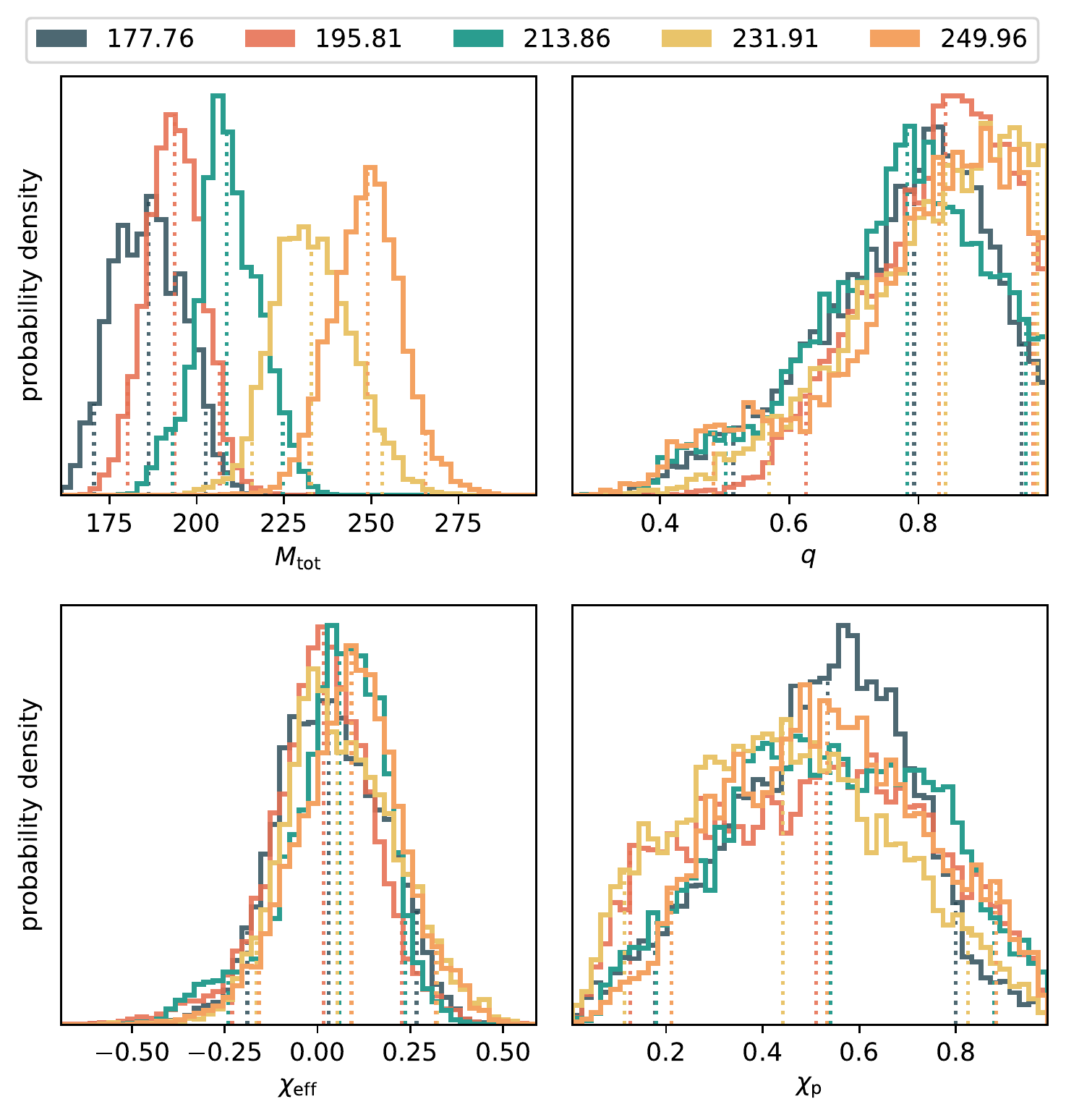}
    \caption{\label{fig:long_wave} Results of injections with different total mass. The lower total mass injection has more cycles in the detectors. The injected total masses are listed in the legend.
    }
\end{figure}

In addition, we also explore lower total-mass binaries, which will have more cycles and duration in the aLIGO-aVirgo band. This means we can extract more information from the signal, potentially increasing the confidence of accurately recovering precession.

To achieve that, we inject signals varying the total mass while keeping other parameters fixed. Since we are investigating black hole binaries in the PISN mass gap and merging to an IMBH, we select 5 masses of primary black holes from 65 $M_\odot$ to the maxL value 98 $M_\odot$ in the source frame. This results in the injected total masses in detector frames being 177.76, 195.81, 213.86, 231.91, and 249.96 solar masses.

The results are shown in Fig. \ref{fig:long_wave}. It shows that the mass ratios recovered from longer waveform injections are more accurate than the shorter ones. 
The longest injected signal (blue line) has the best estimation, with the narrowest posterior, of mass ratios, while the shortest injection is the broadest.
For the $\chi_\mathrm{p}$ recovery, even though the longest injection is still not peaking close to the injected $\chi_\mathrm{p}$, the max likelihood $\chi_\mathrm{p} = 0.75$ is very close to the injected value $\chi_\mathrm{p} = 0.70$.

For lower mass systems that maintain a primary mass within the PISN gap and merge into an IMBH, the results display a more robust indication of precession. However, the precessing SNR remains below 2.1 at a 95\% confidence level, indicating that it is still insufficient to draw a definitive conclusion.

In the following sections, we will continue our exploration of IMBH systems and their detectability in the context of gravitational wave astronomy. By considering GW190521 as a representative example and extending our analysis to a broader range of scenarios and parameters, we aim to gain a more comprehensive understanding of these systems and their implications for our understanding of the universe.

\subsection{Effect of Gaussian Noise on Short-Duration IMBH merger}
\label{sec:res:noise}

In this section, we investigate the influence of Gaussian noise on the parameter estimation of short signals, particularly focusing on the biases it may introduce. We generated 38 different injections using our fiducial waveform embedded in Gaussian noise (referred to as ``noisy injections''), as described in Sec.~\ref{sec:Methodology:injection}. These noisy injections enable us to perform preliminary investigations into the effects of noise on the recovered parameters. 
We expect that detector noise will affect the results even more strongly, so our investigation here provides a lower bound on these effects.

\begin{figure}[htbp]
    \includegraphics[width=0.9\linewidth]{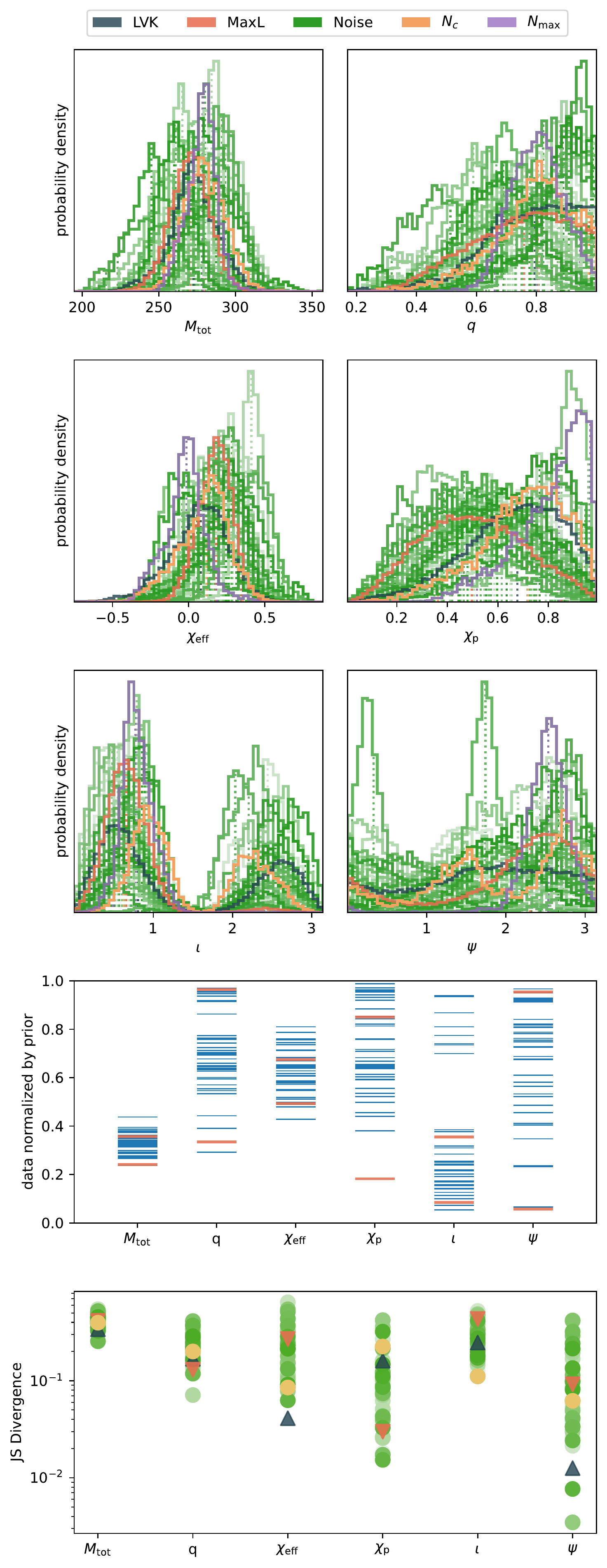}  
    \caption{\label{fig:noise} Results of the noisy injections (green) compared with GW190521 (blue) and the zero-noise maxL injection (red). We highlight the injections which give results closest to GW190521 (orange) and that give the greatest deviation from the prior (lilac). Penultimate panel: maxL values from the noisy injections (blue) compared to the 90\% credible interval of the zero-noise injection (orange). The values are all re-weighted by the range of the priors. Final panel: JS-divergence between the posteriors and the prior.}
\end{figure}

It is clear from these results that noise fluctuations can significantly impact the signal, especially for short signals. The impact is most pronounced for the spin parameters, as evident from the maxL values for $\chi_\mathrm{p}$ and $\chi_\mathrm{eff}$ in some injections that lie outside the 90\% credible interval of the zero-noise injection. 

The total mass and mass ratio are also affected by noise, but to a lesser extent. The addition of noise can shift the peak outside the 90\% confidence interval of the zero-noise injection for the total mass, although the overall shift does not significantly alter the astrophysical conclusions about these types of events. 

The noise-induced changes in the posterior of $\chi_\mathrm{p}$ observed in this analysis are consistent with the findings of Ref.~\cite{Biscoveanu:2021nvg}. However, some differences might arise due to the choice of PSD employed in each analysis. 

From Fig.~\ref{fig:noise}, it is evident that the addition of noise results in deviations of the recovered parameters compared to the zero-noise maxL injections. For the parameters considered here, they do not show systematic biases, but are instead distributed randomly around the zero-noise injection. 

The addition of noise also reveals a second peak in the posteriors for extrinsic parameters such as polarization $\psi$ and inclination $\iota$. This second peak, which was present in the event posteriors but not seen in the zero-noise injection, confirms that this feature arises due to the presence of noise in the signal.

In light of these findings, caution should be exercised when inferring the nature of events characterized by short signals based on the location of the peak and the values of the maxL parameters. Real-world non-Gaussian noise could introduce additional biases, emphasizing the importance of considering the influence of noise when interpreting results.

\subsection{Other affecting factors}

\subsubsection{Noise on Non-Spinning Systems}
\label{sec:results:zero_spin_noise} 

\begin{figure}[t]
    \includegraphics[width=\linewidth]{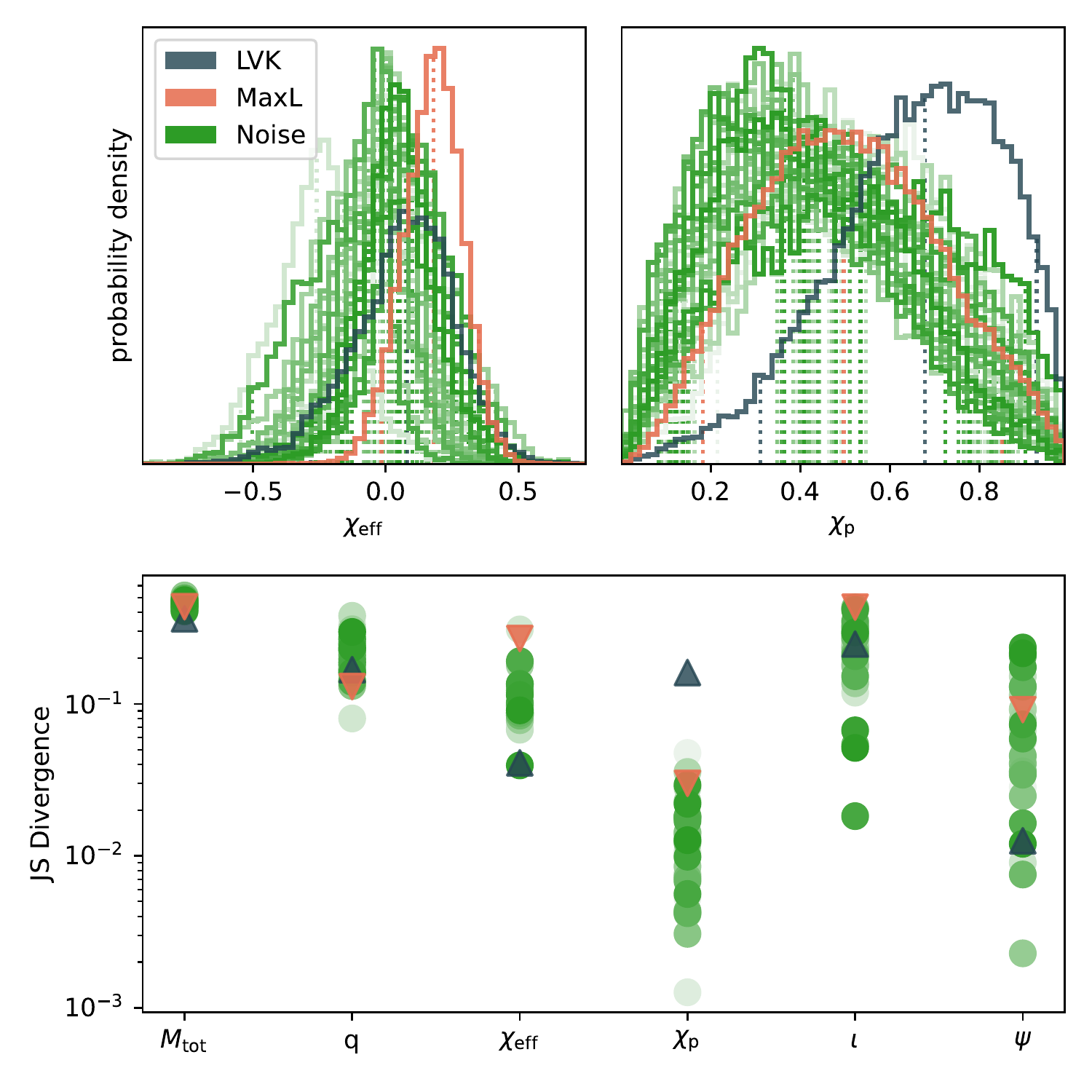}  
    \caption{\label{fig:noise_nonspinning} Results of non-spinning noise injection. The injected signals are generated with the same configuration as Fig.~\ref{fig:noise} except the spin components are set to zero. The dots in the bottom panel represent the JS divergence from posteriors with respect to the prior.
}
\end{figure}

In the previous section, we found that a combination of Gaussian noise and a highly precessing signal is sufficient to show indications of high in-plane spins. This motivates us to investigate whether the evidence for high precession can be mimicked purely by Gaussian noise from a non-spinning or aligned-spin binary. To do so, we generated another set of injections using 30 random noise realizations. The injected signals have the same parameters as in Tab.~\ref{tb:maxL}, but with spin components set to zero. We chose a non-spinning binary as the limiting case.

The results of these injections are shown in Fig.~\ref{fig:noise_nonspinning}. We find that the intrinsic parameters are less affected by the addition of noise than in the precessing case, as can be seen by comparison of the JS divergence. As expected, the posteriors for $\chi_\mathrm{eff}$ are centred on zero (the injected value) and do not appear to be noticeably impacted by the noise. 
This is also clear for the posteriors of $\chi_\mathrm{p}$, where we essentially recover the prior for each of our 30 injections. It is clearly unlikely that evidence of high precession seen in the LVK results can be reproduced by pure noise. The JS plot corroborates the fact that the noise cannot cause a sufficient deviation from the prior to mimic a highly precessing system. It is therefore not possible to produce evidence for precession in such a system purely as a result of Gaussian noise

\subsubsection{Effects of Varying Spin Magnitude}

Since our results indicate that a non-spinning binary cannot give indications of precession even in the presence of Gaussian noise, we aim to determine the minimum in-plane spin magnitude required to obtain precession information for our fiducal IMBH source.

The specific spin configuration significantly impacts the recovery of $\chi_\mathrm{p}$ from a short, highly precessing signal~\cite{Biscoveanu:2021nvg}.
For certain configurations, there is a much higher probability of observing evidence of precession in the signal.
However, as demonstrated, this is not the sole effect that can produce evidence of a highly precessing system.
Since we do not intend to investigate the effect of spin orientation here, we fixed the orientation of the spin to the value given in Tab.~\ref{tb:spin}.

\begin{table}[t]
\caption{\label{tb:spin} Injected spins of varying magnitude.}
\begin{ruledtabular}
\begin{tabular}{l|rrrr}
  & $\chi_\mathrm{p}$ & $\chi_\mathrm{eff}$ & $a_1$ & $a_2$\\
  \colrule
  $\chi_{p1}$& 0.15 & 0.02 & 0.020 & 0.20\\
  $\chi_{p2}$& 0.30 & 0.04 & 0.040 & 0.39\\
  $\chi_{p3}$& 0.46 & 0.06 & 0.059 & 0.59\\
  $\chi_{p4}$& 0.62 & 0.08 & 0.079 & 0.79\\
  $\chi_{p5}$& 0.77 & 0.10 & 0.099 & 0.98\\
  maxL &   0.70& 0.09 & 0.090 & 0.90
\end{tabular}
\end{ruledtabular}
\end{table}

\begin{figure}[t]
    \centering
    \includegraphics[width=\linewidth]{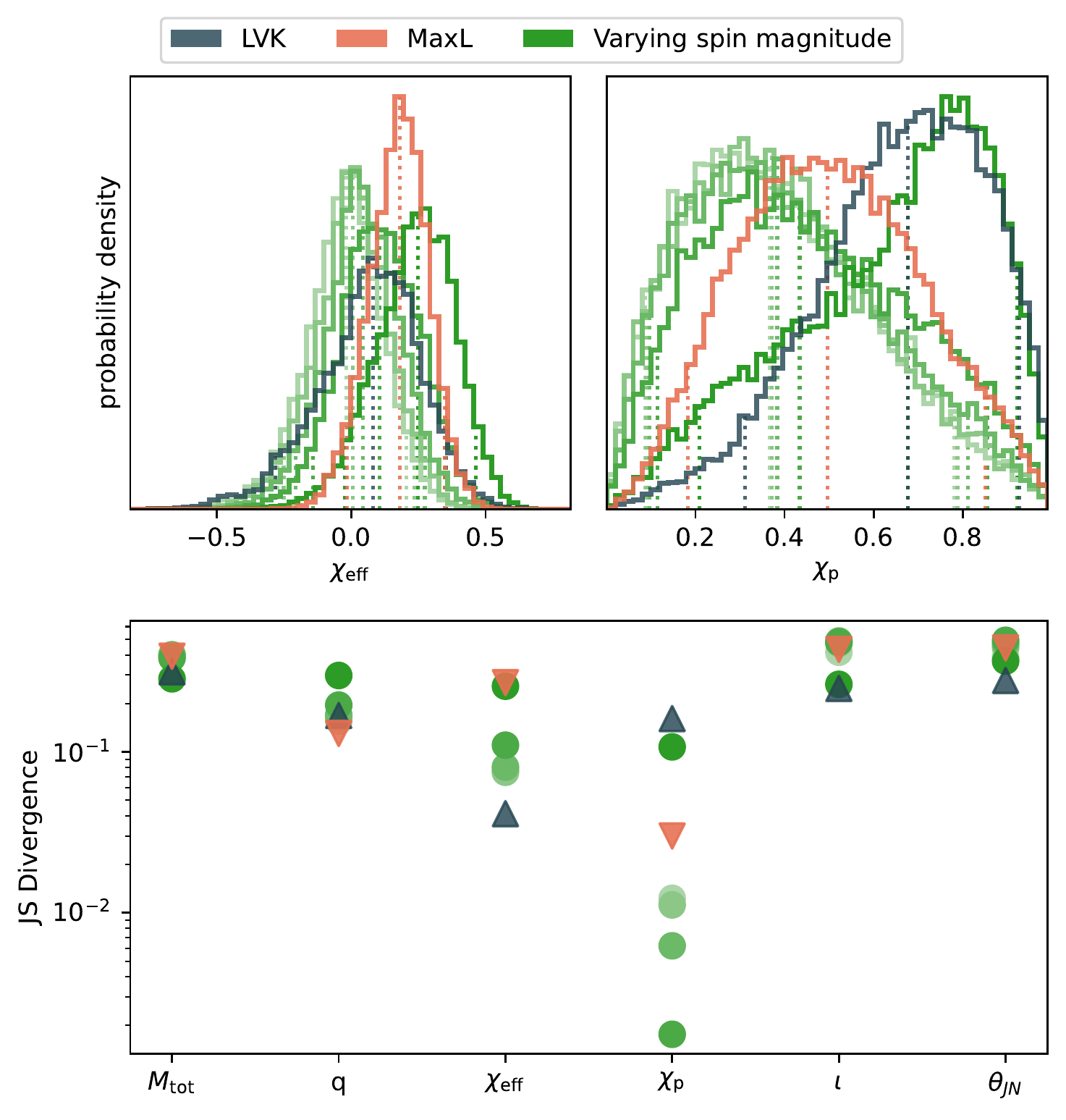}
    \caption{\label{fig:spin_mag} Results of the injections with varying spin magnitude in zero noise. The green lines are the injections with different spin magnitudes, which are listed in Tab.~\ref{tb:spin}.
    These are compared with the results from GW190521 (blue) and the zero-noise maxL injection (orange).
    The bottom panel give the JS-divergence from posteriors to prior.}
\end{figure}

As shown in Fig.~\ref{fig:spin_mag}, deviations from the prior in $\chi_\mathrm{p}$ start to appear when $\chi_\mathrm{p}$ is around 0.7, and become significant when $\chi_\mathrm{p} \sim 0.8$. The JS divergence between the posteriors and the prior for $\chi_\mathrm{p}$ supports this conclusion, with only the $\chi_{p5}$ injection showing a deviation of $\mathcal{O}\left(10^{-1}\right)$ bits. For a zero-noise injection, an in-plane spin of about $0.77$ and a total spin greater than $0.9$ (beyond the calibration range of NRSur7dq4) is necessary to produce a deviation from the prior that indicates high in-plane spins.

This investigation can help inform our interpretation of the GW190521 event and contribute to our astrophysical understanding of the IMBH binary population. Notably, the JS divergence between $\chi_{p5}$ and the LVK results is only 0.014 bits, which means the posteriors for $\chi_\mathrm{p}$ are quite similar to those observed for the event. However, we have not yet considered the effects of noise on the signal, potential degeneracy with eccentricity, or the impact of extrinsic parameters. We will reevaluate any conclusions drawn from this analysis in the following sections.

\begin{figure}[ht]
    \centering
    \includegraphics[width=\linewidth]{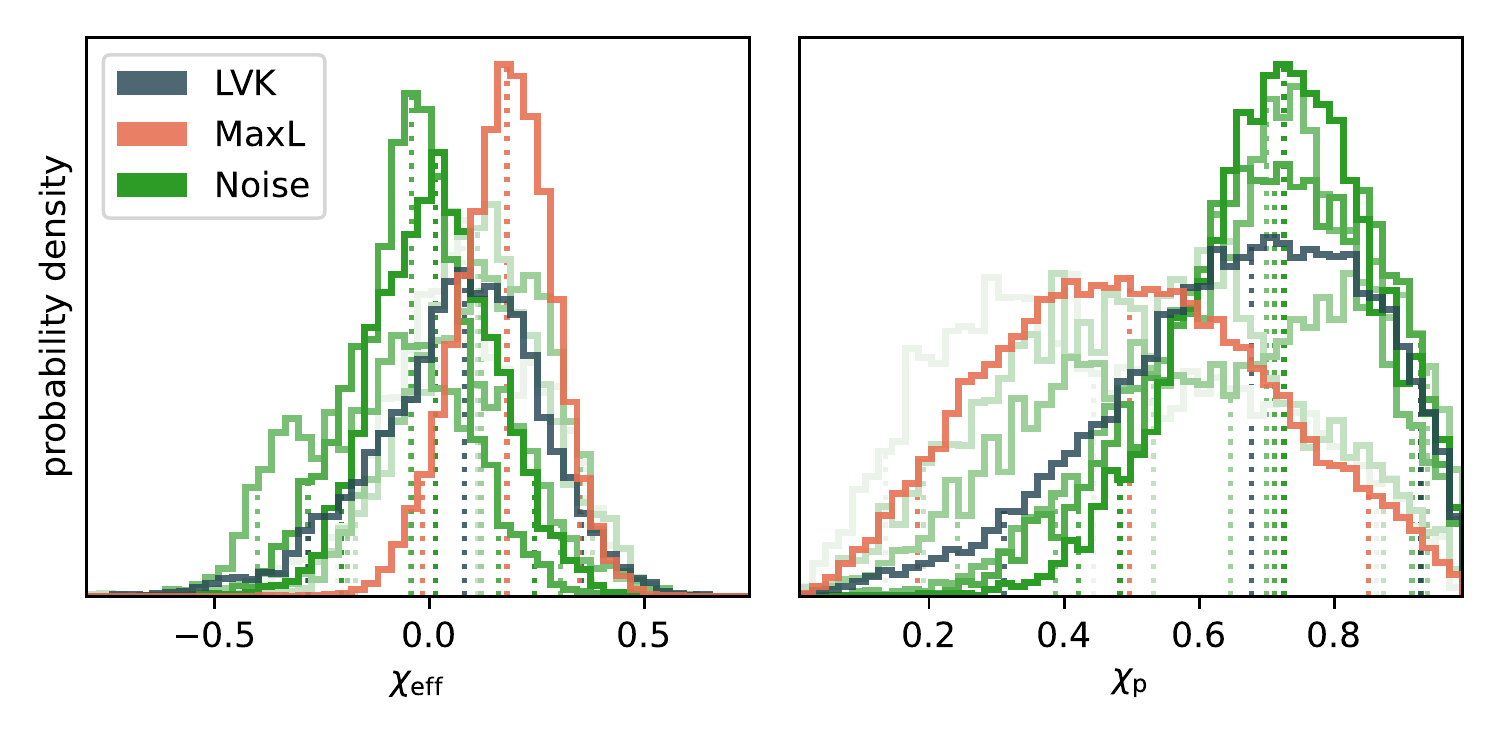} 
    \caption{\label{fig:spin_mag_noise} Results of the injections varying spin magnitude into the $N_\mathrm{max}$ noise realisation. The green lines are the injections with different spin magnitudes, which are listed in Tab.~\ref{tb:spin}.
    These are compared with the results from GW190521 (blue) and the zero-noise maxL injection (orange).}
\end{figure}

To investigate the minimum spin magnitude required when accounting for detector noise, we injected our signal into the $N_\mathrm{max}$ noise realization, which produced the largest inferred value of $\chi_\textrm{p}$ as discussed in Section~\ref{sec:res:noise}. The results are presented in Fig.~\ref{fig:spin_mag_noise}. We find that in the presence of noise, a value of $\chi_\mathrm{p}$ greater than 0.4 is necessary to provide evidence of high in-plane spin values. A systematic exploration using a range of noise realizations has not been conducted due to computational resource limitations. However, combined with the results in Section \ref{sec:results:zero_spin_noise}, this analysis shows that a high inferred effective precession spin for this type of event requires both a non-zero value of $\chi_\textrm{p}$ (indicating some degree of precession in the source signal) and noise. The presence of noise in the signal considerably reduces the magnitude of the spin that can be claimed for this event.

\subsubsection{Extrinsic Parameters and Their Impact} \label{sec:results:extrinsic}

The measurability of precession is affected by the extrinsic parameters, especially polarization $\psi$ and inclination $\iota$ \cite{Green2020}. 
These parameters are poorly measured from the signal. We examine whether we could confidently detect precession from a highly precessing, equal mass IMBH by choosing the values of $\iota$ and $\psi$ that maximize the likelihood of detecting any precession present in the signal.

\begin{figure}[t]
    \includegraphics[width=0.48\textwidth]{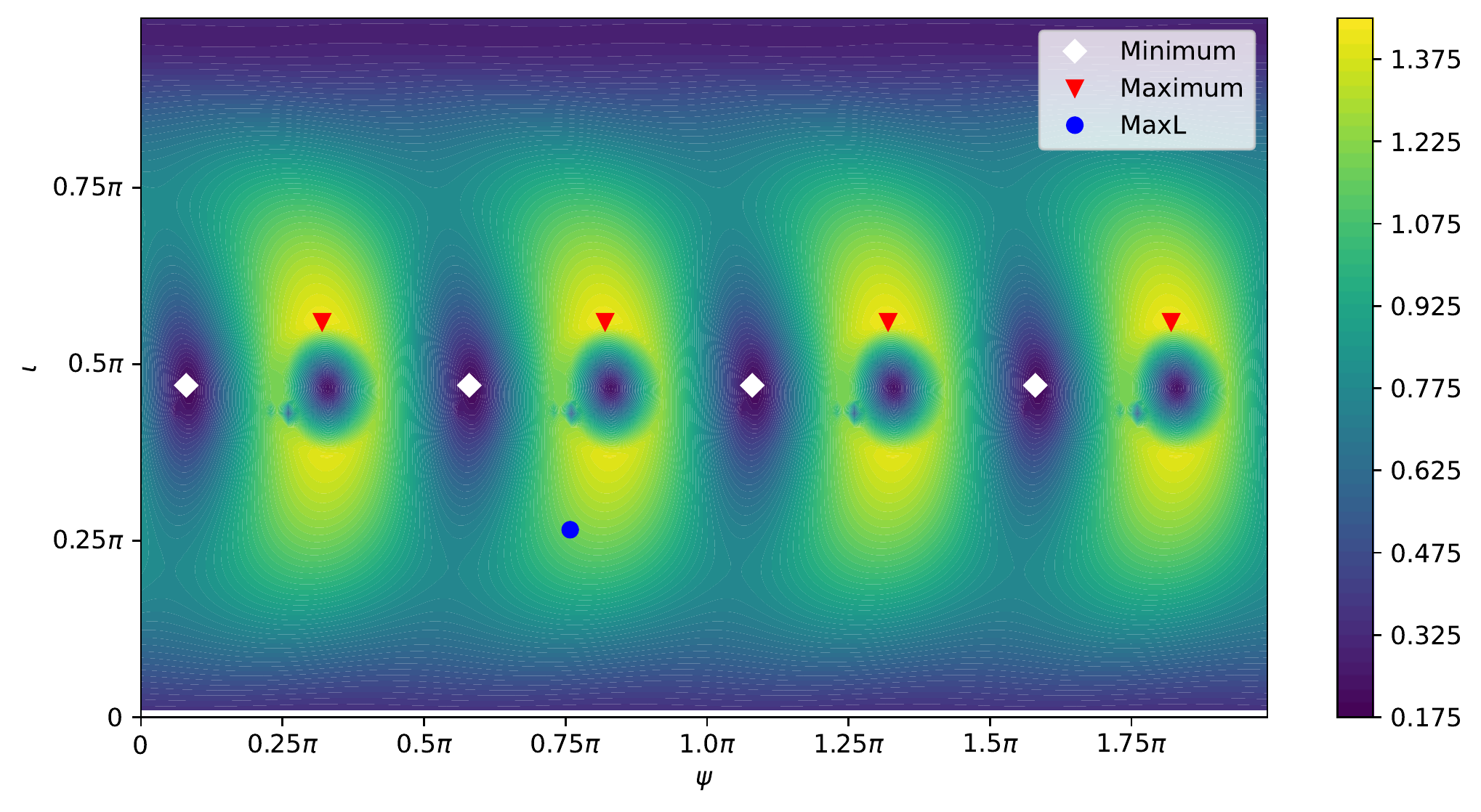}
	\caption{\label{fig:psi_iota} Calculation of $\rho_\mathrm{p}$ with varying inclination $\iota$ and polarization $\psi$ using maxL parameters. The color bar on the right displays the value of $\rho_\mathrm{p}$. The blue circular dots in this figure represent the maxL value, the white triangular dots are the minimum values, and the red diamond dots are the maximum values.}
\end{figure}

We calculated $\iota$ and $\psi$ for which the precessing SNR $\rho_\mathrm{p}$ is maximised for a binary with the properties listed in Tab.~\ref{tb:maxL}. The dependency of $\rho_\textrm{p}$ on $\psi$ and $\iota$ is shown in Fig.~\ref{fig:psi_iota}. 
From this, we can see that our choice of $\iota$ and $\psi$ for our example system lie in a region where, compared to the maximum possible value of $\rho_\mathrm{p}$ for this event, we are only moderately confident in our ability to detect precession. 
The maximum value of $\rho_\mathrm{p}$ possible for such a system is around 1.43, which is still below the threshold to unambiguously claim detection of precession, as discussed in Sec.~\ref{sec:Methodology:psnr}. We therefore do not expect to be able to claim a detection of precession for any set of extrinsic parameters for this event

Nonetheless, we aim to examine the influence of these extrinsic parameters on our ability to interpret the signal. We generated and injected another waveform based on the parameters in Table~\ref{tb:maxL}, altering the values of $\iota$ and $\psi$ to maximize $\rho_\mathrm{p}$ to investigate how optimizing these extrinsic parameters impacts the likelihood of detecting precession for a given system. We chose the first maximum in Fig.~\ref{fig:psi_iota} ($\iota = 1.759$ and $\psi=1.005$) since the four maxima are equivalent. We also investigated the effect of noise on these modified extrinsic parameters, using the $N_{\mathrm{max}}$ noise realization.

\begin{figure}[t]
    \includegraphics[width=\linewidth]{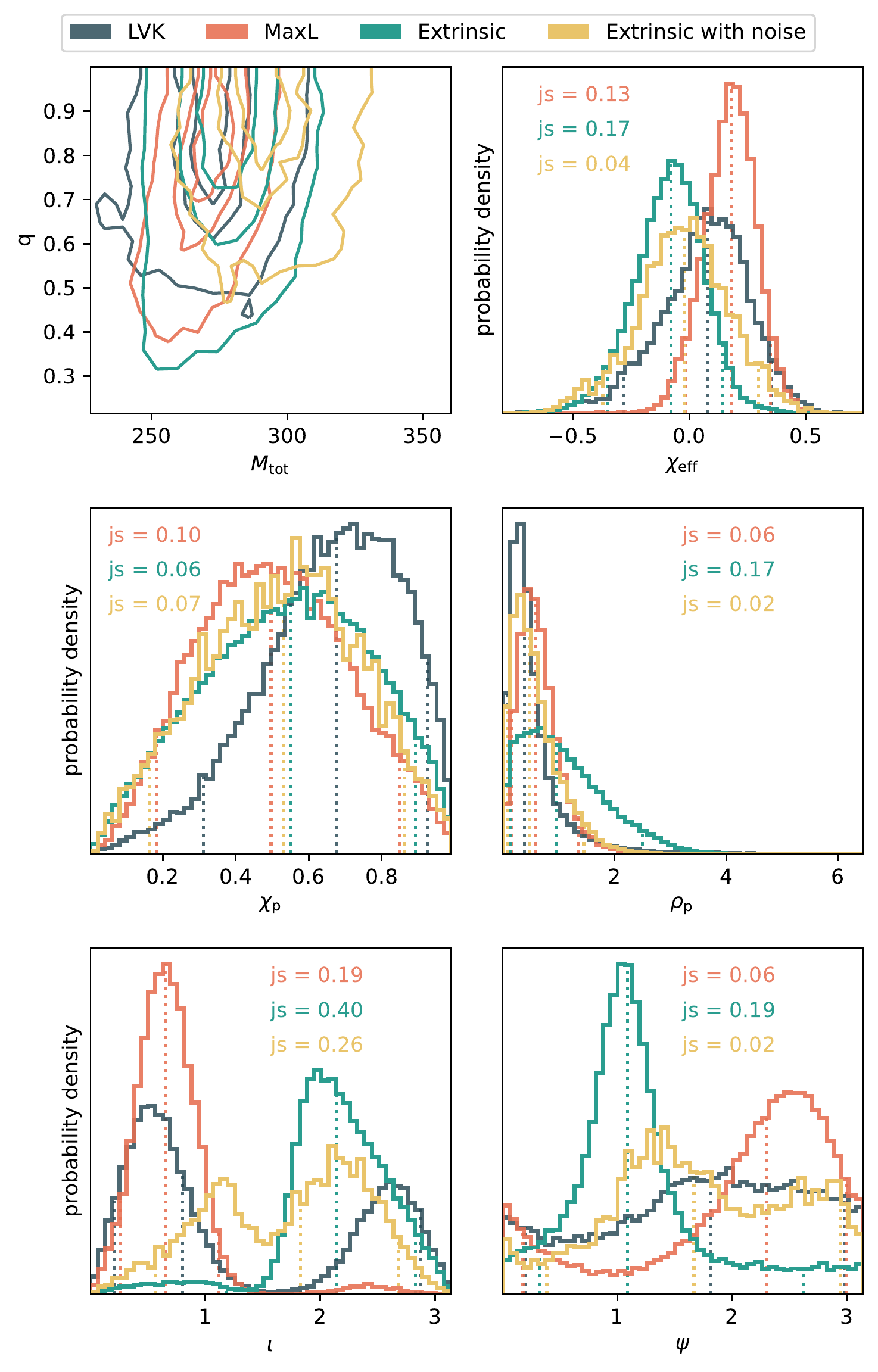}   
    \caption{\label{fig:max_rho_p} Results of injections using optimized extrinsic parameters. The green line represents the zero-noise injection with $\iota = 1.759$ and $\psi=1.005$, the values that maximize the precessing SNR. The yellow line represents a noisy injection with the same extrinsic parameters and the $N_\mathrm{max}$ noise realization.}
\end{figure}

From the results (shown in Fig.~\ref{fig:max_rho_p}), the peak of the posteriors for the inclination $\iota$ and polarisation $\psi$ in the zero-noise max $\rho_\mathrm{p}$ injection is centred on the injected value.

Similar to our original case, we observe only a single peak for each of these parameters in the absence of noise. With the addition of noise (using the $N_\mathrm{max}$ realization), the single-peak distributions disappear, and we see posteriors that closely resemble the LVK results.

We also see that when injecting the parameters that maximize the precessing SNR, the peak of the posterior for $\chi_\textrm{p}$ shifts to slightly higher values. However, it is still insufficient to claim a detection of precession. Interestingly, unlike the case with sub-optimal values of $\iota$ and $\psi$, the addition of noise does not significantly affect the distribution for $\chi_\textrm{p}$, so even with noise, we cannot recreate evidence for high in-plane spins for a binary at this sky location. This is consistent with our expectation since $\rho_\textrm{p}$ remains below the threshold at which we expect to measure precession. Regardless of the true values of $\psi$ and $\iota$, we still would not be able to detect precession for an equal mass, highly precessing binary. However, the apparent evidence of precession in the signal may enable us to rule out certain sky positions.

\section{Investigation of eccentricity}\label{sec:ecc}

It has been suggested that for this kind of system, evidence of precession in the signal may degenerate with indications of eccentricity~\cite{Romero-Shaw:2020thy, Gayathri2022, Romero-Shaw:2022fbf}. 
To examine this possibility, we investigate the likelihood of seeing evidence for precession in the signal from a heavy 
eccentric binary when recovering such a signal with a non-eccentric waveform model. 
We use a selection of NR waveforms taken from the SXS and RIT catalogues as discussed in Sec.~\ref{sec:metho:wf}.

We first performed a systematic investigation of low eccentricity ($e<0.2$) systems. This is motivated in part by the public availability of NR waveforms of sufficient length to perform this study. Nonetheless, we expect the results of this study to be of general interest since most astrophysical models predict the majority of binaries have $e<0.2$ by the time they enter the LIGO band ($> 20 \mathrm{Hz}$) \cite{Samsing:2017xmd, Gondan:2017wzd, Antonini:2017ash}.

However, some astrophysical models \cite{Tagawa:2020jnc,Silsbee:2016djf,Liu:2018vzk} 
do suggest binaries with much higher eccentricities are possible --- that would make this a very rare event and provide much information about the formation channel of the binary. Therefore, we also performed a limited study of highly eccentric systems.

Before using any NR waveforms in our analysis, we need to have a consistent method of estimating eccentricity from these waveforms where possible, as will be outlined in Appendix ~\ref{sec:eccentricity estimator}.
We then perform an injection study with these waveforms, following the procedure detailed in Sec.~\ref{sec:Methodology:injection}.
The results of this study are given in Sec.~\ref{sec:ecc:injection}.

\begin{figure}[htbp]
    \includegraphics[width=\linewidth]{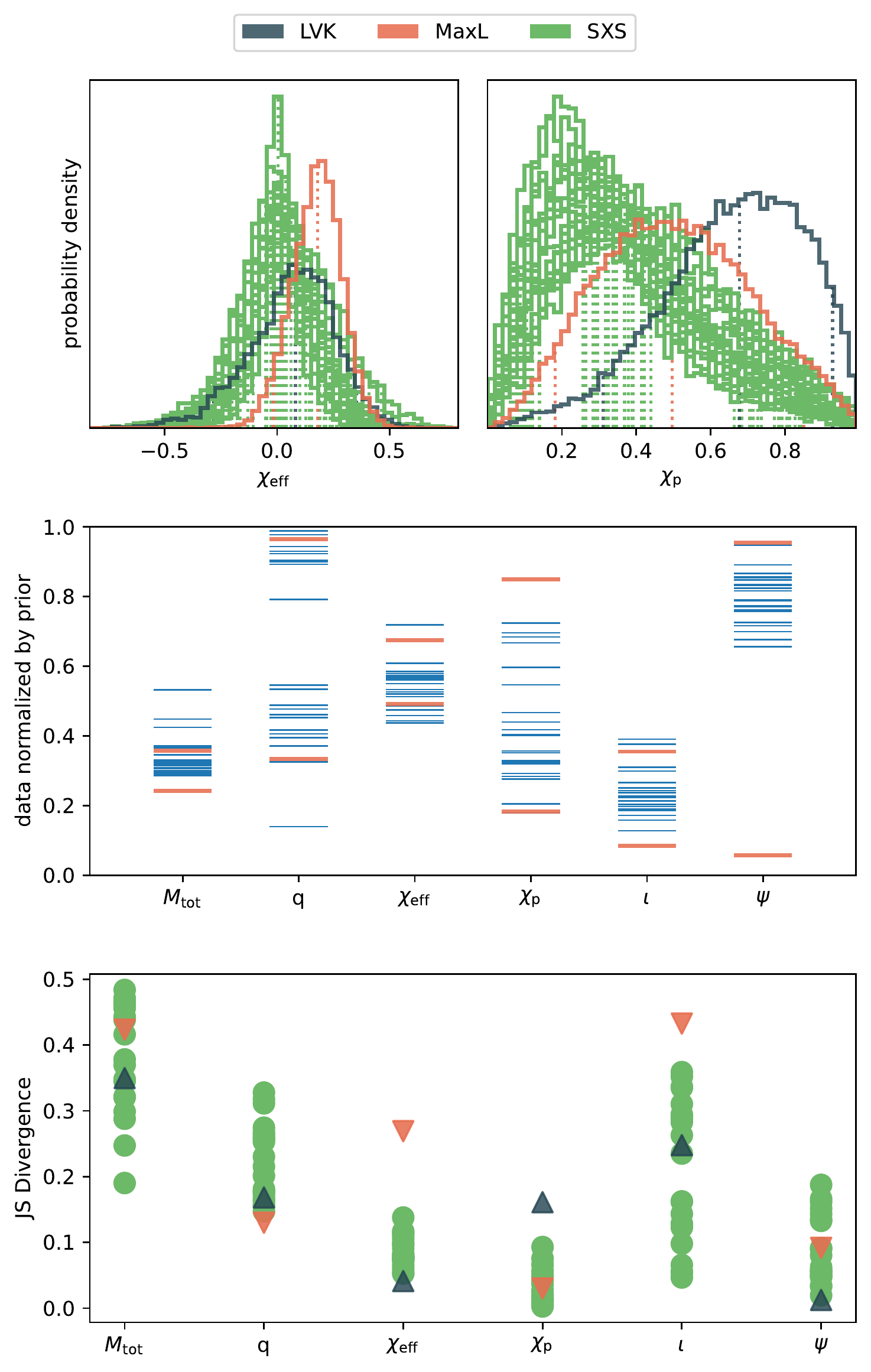}
    \caption{\label{fig:sxs} The results of the SXS injections. The posteriors from the eccentric injections (green) are compared with the results from GW190521 (blue) and the zero-noise maxL injection (orange). The penultimate panel shows the maxL values of the eccentric injections (blue) compared to the 90\% confidence interval of the zero-noise injection (orange). The bottom panel shows the JS divergence between the posteriors and the prior.
    }
\end{figure}

We use the Numerical Relativity Injection Infrastructure \cite{Schmidt:2017btt} to generate a waveform with the same total mass and extrinsic parameters shown in Tab.~\ref{tb:maxL}. 
We then perform the injection as detailed in Sec.~\ref{sec:Methodology:injection}.

Since the eccentricity decays with time, we need to choose a reference time $t_\mathrm{ref}$ at which to report the eccentricity. Here we choose $t_\mathrm{ref} = -2500 M$ because this is the earliest consistent time among all the waveforms taken from the SXS catalogue.
The $t=0$ is defined as the merger time, which is the peak of the strain. The estimated eccentricities are given in Tab.~\ref{tb:sxs}.

For the cases taken from the RIT catalog, 
we note that the waveforms only contain 0.5 cycles prior to the merger. 
All the methods proposed to estimate the eccentricity from the waveform \cite{Buonanno:2006ui, Husa:2007rh, Mroue:2010re, Ramos-Buades:2019uvh} are not valid for such short waveforms.
Consequently, we have to rely on the eccentricity reported in the metadata of each waveform~\cite{Healy2022wdn} and cannot perform an independent estimate. 
Thus, we cannot ensure the consistency of eccentricities with those we report for the SXS waveforms. 
To distinguish the two estimates, we label the eccentricity taken directly from the metadata as $e_r$.

\label{sec:ecc:injection}
\subsection{Results}

The results of the systematic investigation into the impact of comparatively low eccentricity on the quasi-circular PE of a high mass system are shown in Fig.~\ref{fig:sxs}.
The median and 90\% distributions of the key parameters from the posterior are also given in Tab.~\ref{tb:sxs}. 
From the $\chi_\mathrm{p}$ plot, we can see that 
for a signal at SNR 15.4 from a high mass non-spinning binary with $q \le 3$ and $e \le 0.2$, the effect of eccentricity on the signal does not mimic the effect of precession.
The JS divergence of the $\chi_\mathrm{p}$ posteriors from the prior is $\mathcal{O}\left(10^{-2}\right)$ bits, implying that the PE is simply recovering the prior in these cases.
From this, we can infer that the evidence of high-precessing spins cannot be produced by eccentricity for short signal, heavy mass binaries with $e<0.2$. 

In Ref.~\cite{Gayathri2022}, it was proposed that GW190521 originated from a highly eccentric black hole merger with $e = 0.69^{+0.17}_{-0.22}$. Here, we consider two waveforms, the first a non-spinning waveform (RIT:eBBH:1318) and the second with $\chi_\mathrm{p} = 0.7$ (RIT:eBBH:1639), both with the initial eccentricity $e_r=0.75$.
The precessing waveform was found in~\cite{Gayathri2022} to be that which most closely matched the posteriors from GW190521. 
We performed both a zero-noise and noisy injection using the $N_\textrm{max}$ noise realisation of each of these waveforms.

\begin{figure}[t]
    \includegraphics[width=\linewidth]{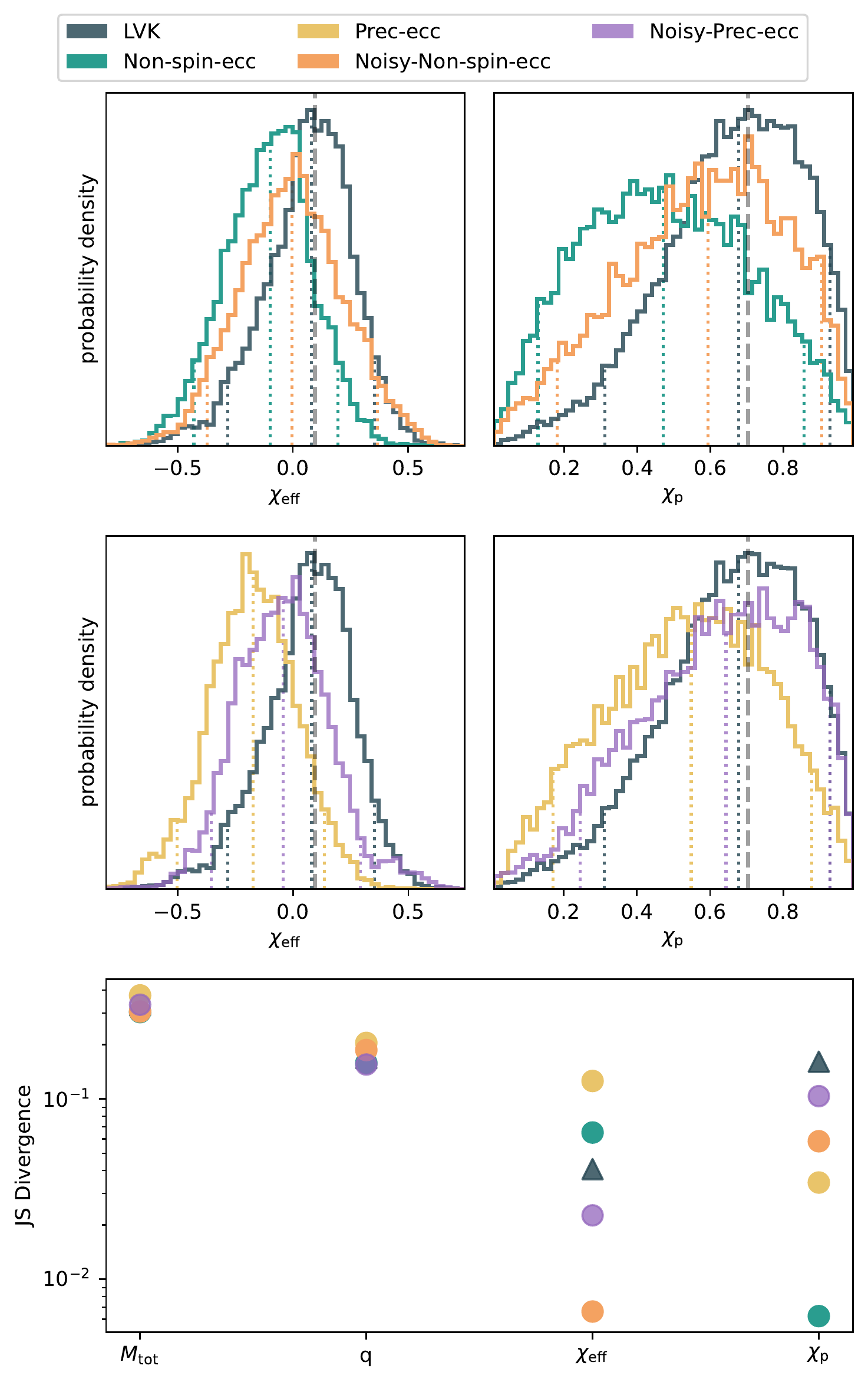}
    \caption{\label{fig:rit} Results from the high eccentricity injections. In zero-noise, neither the non-spinning eccentric waveform (green) nor the precessing eccentric waveform (yellow) shows evidence of high in-plane spins. This is changed with the addition of noise, though the non-spinning waveform (orange) does not indicate such high spins as the precessing waveform (purple). The bottom panel shows the JS divergence between the posteriors and the prior.
    }
\end{figure}

The results are shown in Fig.~\ref{fig:rit} and the median and 90\% distributions of the key parameters from the posterior are given in Tab.~\ref{tb:rit}. 
We can see that the non-spinning eccentric waveform does not cause significant deviation from the prior for $\chi_\mathrm{p}$, with only JS divergence $D_\mathrm{JS} = 0.006$ bits. It shows no difference to the non-spinning non-eccentric case.
The injected signal with both a high degree of precession and eccentricity shows a greater deviation from the prior --- with a JS divergence of 0.034 bits. 
However, this is similar to the amount of information about the degree of precession in the system that we are able to infer from a high precession, quasi-circular signal (which has a JS divergence of 0.029 bits with respect to the prior). It is therefore difficult to identify whether the indication of precession in these results is due to the high in-plane spins or the high eccentricity. In addition, the combination of eccentricity and precession is still insufficient to reproduce the very high degree of precession seen in the posteriors for GW190521.

Adding noise to the signal causes both the zero-spin and the precessing cases to show significant deviation from the prior (0.058 and 0.104 bits respectively). As in the non-eccentric case, noise appears to be a key requirement in creating apparent evidence for high in-plane spins. 
Both the precessing and the non-spinning signals are now close to now show significant deviations from the prior with a peak between 0.5 and 0.7 (with the non-spinning case peaking at lower values).
For the precessing signal, this is as was seen in the quasi-circular case.
For the non-spinning system, this deviation from the prior is specific to the eccentric case.
In the presence of noise, it is therefore possible to produce evidence for a precessing system from a non-spinning, eccentric binary.
There is therefore no clear evidence that this signal is any more likely to have come from an eccentric binary than from a quasi-circular one.

We therefore do not see any degeneracy between eccentricity and precession for this kind of high-mass system in the absence of noise.
In the presence of noise, either precession or high eccentricity (or a combination of both) can give the appearance of a highly-precessing signal.
It is not, however, possible to identify the cause.

\section{Higher SNR and future detectors for heavy mass binaries} 
\label{sec:res:future_det}

\begin{figure}[t]
    \includegraphics[width=\linewidth]{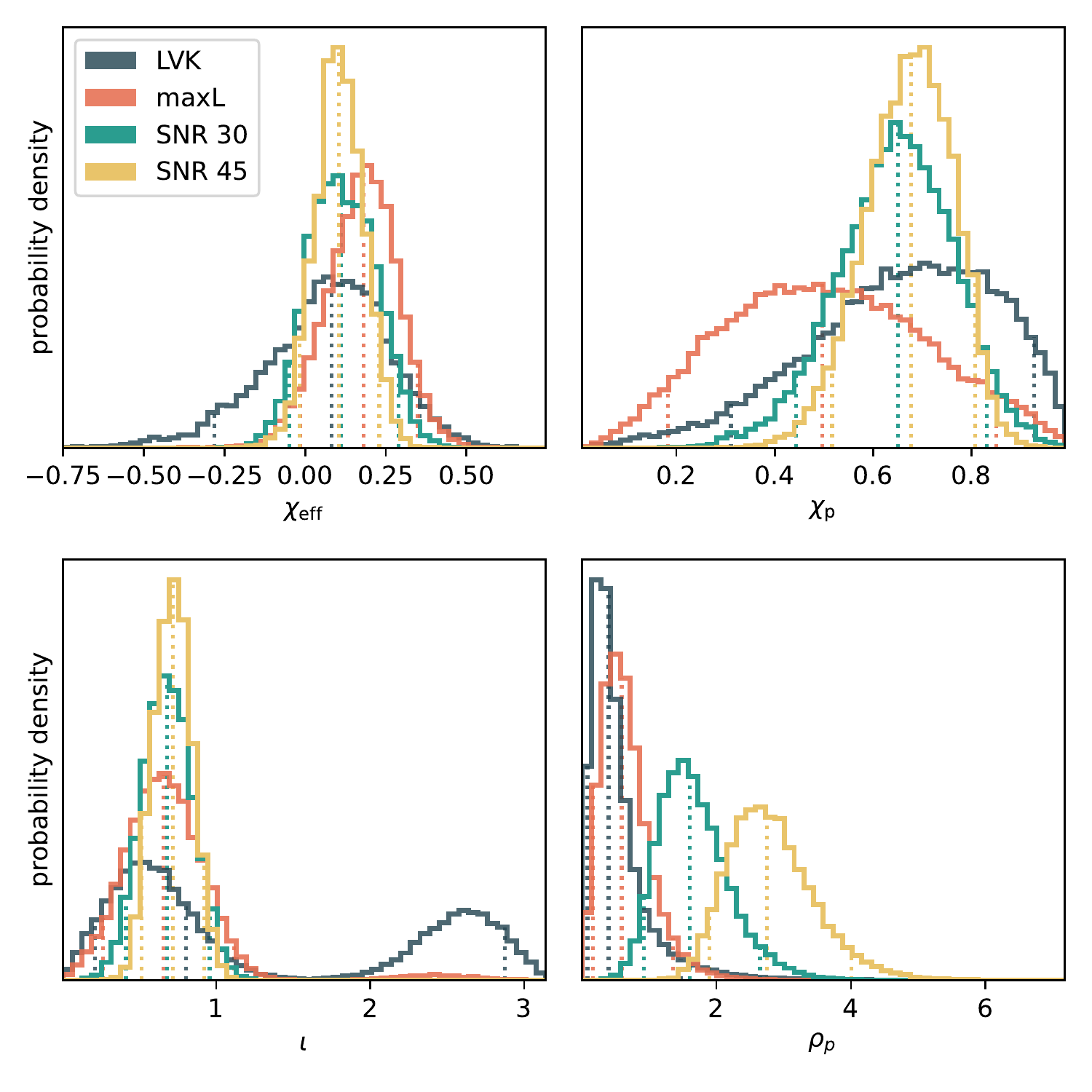}
    \caption{\label{fig:snr} Results of injections with varying SNR. The SNR for the maxL injection is 15. We show an injection of SNR 30 (green) and SNR 45 (yellow). Only the injection with SNR 45 has $\rho_\mathrm{p}$ at 90\% confidence, implying that only above this SNR can we unambiguously claim a detection of precession.}
\end{figure}

Besides, we want to explore under what conditions we will be able to confidently detect precession from a highly precessing system with high total mass and a mass ratio close to 1. We examined two possibilities: our example binary at a closer distance, and thus with a higher SNR for an aLIGO and aVirgo network;  and the same source at the same distance but using the PSD from future third-generation (3G) detectors \cite{sensitivity_curve} 
which will be more sensitive and thus the detection will have a higher SNR.

To explore how high an SNR is needed for an aLIGO and aVirgo network to detect precession, we performed two injections with the maxL parameters but with varying SNRs (SNR 30 and SNR 45). Fig.~\ref{fig:snr} shows that in order to be able to confidently detect precession (i.e. for $\rho_\mathrm{p} > 2.1$) for such a high total mass, low mass ratio, highly precessing signal (at 95\% confidence), the SNR of the signal needs to be greater than 45,
and possibly even higher, if we increase our threshold value of $\rho_\mathrm{p}$ in accordance with the findings of Ref.~\cite{Pratten:2020igi}.

\begin{figure}[t]
    \includegraphics[width=\linewidth]{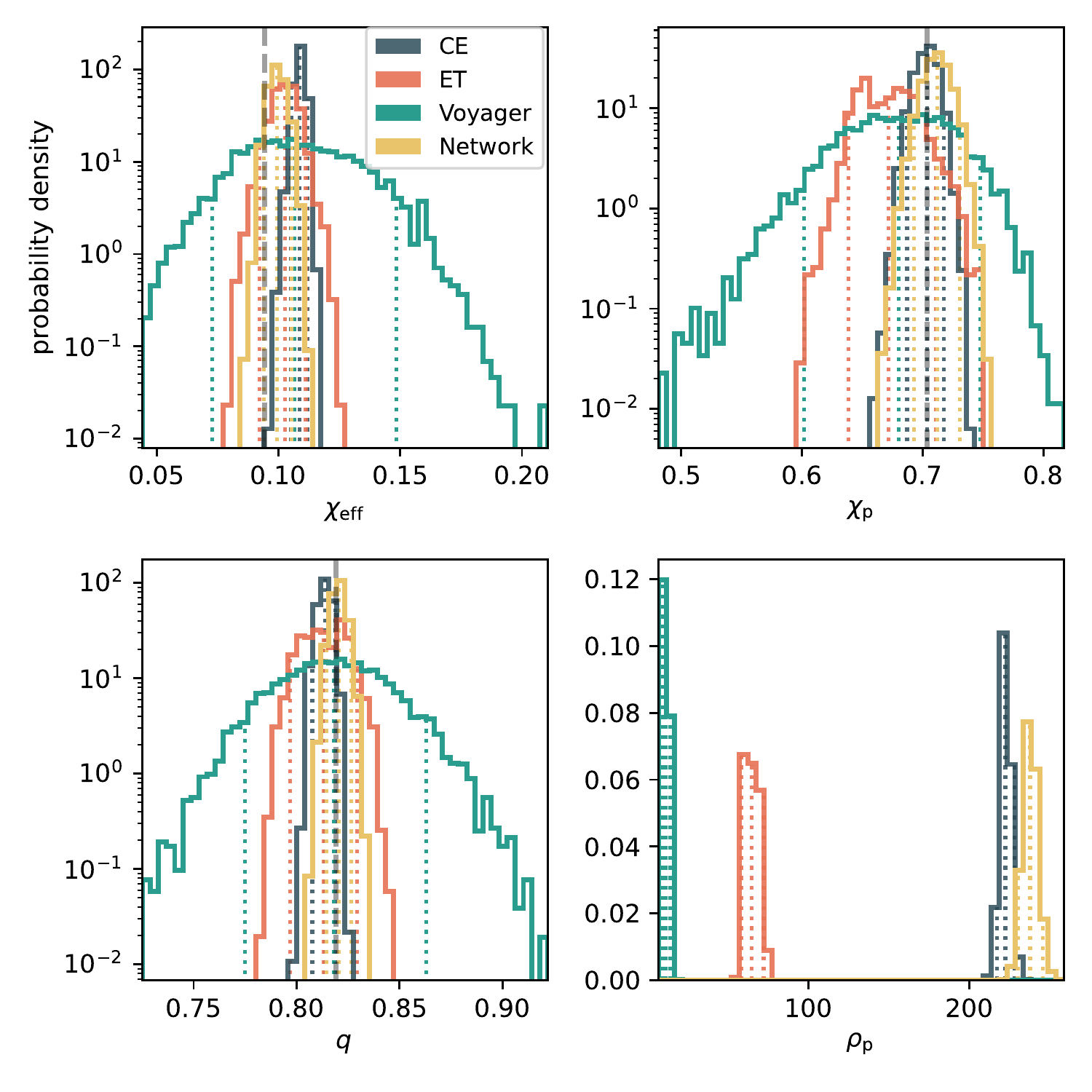}
    \caption{\label{fig:detector} Results of the injections with CE, ET, Voyager and a detector network formed by CE and ET (labelled Network). The grey dashed vertical lines are the injected values. The y-axes in the first three panels are plotted as log scale to show the 90\% credible interval more clearly.
    }
\end{figure}

We considered the increase in sensitivity provided by both individual 3G detectors (using the PSDs shown in Fig.~\ref{fig:sen_curv}) and by a theoretical detector network.
This detector network is formed by ``ET'' (a single interferometer in the position of V1) and ``CE'' (a single interferometer in the position of H1), using the detector response of V1 and H1 respectively.
The only change we consider is an improvement in sensitivity and thus use the appropriate 3G PSD. 
We assume the detectors to have the same configuration and orientation as current ground-based detectors and ignore other properties of 3G detectors such as the triangle configuration.
The results from the PE are shown in Fig.~\ref{fig:detector}.

With the increase in sensitivity of the detectors, the injected signals are much louder and the posterior distributions are consequently narrower.
For each of the 3G detectors, all of the parameters of the injected signal are well recovered, even for a signal of such a short duration.
For a few parameters, the distributions are now sufficiently narrow that the injected value, while very close to the peak of the distribution, lies outside or on the edge of the 90\% credible interval. 
Two examples of this are the recovery of $\chi_\mathrm{eff}$ with CE and $\chi_\mathrm{p}$ with ET.
From this, we can see that we will require more precise waveform models or PE for dealing with such high SNR events.

\section{Discussion}\label{sec:discussion}

In this work, we have investigated the detectability of precession and eccentricity for gravitational wave events within the IMBH mass range, focusing on the challenges presented by short signals with only a few cycles in band. To explore this, we performed an injection study using the NR surrogate model NRSur7dq4 and a selection of NR waveforms taken from the SXS and RIT catalogues. We assessed the degree to which we could measure precession and its degeneracy with eccentricity.

We first injected a signal with the maxL parameters extracted from the detection of our fiducial binary GW190521 and analyzed our ability to recover various parameters. We then extended our analysis to different total masses within the IMBH range. We found that the total mass, mass ratio, and $\chi_\mathrm{eff}$ are generally well recovered. However, in the fiducial parameter injection and most cases involving injections with varying total masses, we found that we essentially just recover the prior for $\chi_\textrm{p}$, giving us limited information about the degree of precession in the injected signal. Since we injected a relatively highly-precessing binary, we therefore investigated other possible causes of the high degree of precession seen in the results for GW190521.

One possible cause we explored was the effect of noise on the signal. We found no evidence of systematic bias in the recovery of the parameters due to Gaussian noise. However, the maxL values of $\chi_\mathrm{eff}$ and $\chi_\mathrm{p}$ of some of the injections lie outside of the 90\% credible interval of the zero-noise injection. Our results show that noise strongly affects the recovery of some parameters for a signal of this length, and we should be cautious of making strong claims about such a short signal. While a highly precessing signal with noise could be responsible for the suggestion of a high degree of precession in the signal, we found that noise alone appears insufficient to mimic high precession. 
We require $\chi_\mathrm{p}>0.75$ for a zero-noise injection, while only $\chi_\mathrm{p}>0.4$ is needed in a given noise realization for replicate the high precession in posterior.
This implies that some degree of precession or eccentricity is required to reproduce these results, assuming the signal to come from a BBH.

The extrinsic parameters of the binary do not significantly affect our conclusions, although optimizing the values of $\iota$ and $\psi$ can increase the chance of detecting precession. Interestingly, for these optimized values, the addition of noise does not have a noticeable impact on the recovery of the intrinsic parameters, unlike with generic values of $\iota$ and $\psi$.

Our investigation of the degeneracy of eccentricity with precession for this event reveals no degeneracy between spin and eccentricity for a low eccentricity ($e \leq 0.2$) system at such high total mass that we see only merger-ringdown. 
This is also the case for the highly eccentric waveforms taken from the RIT catalogue; for the non-spinning eccentric case we only recover the prior for $\chi_\mathrm{p}$, while the highly precessing eccentric injection is still insufficient to reproduce LVK the results seen for GW190521 and it is impossible to determine whether the slight deviation from the prior is due to the effect of precession or eccentricity on the signal. 
Similar to the quasi-circular case, the addition of noise makes it possible to reproduce evidence of highly precessing spins in the signal, for both the non-spinning and the precessing injections.
It is consequently difficult to conclusively state which of these effects could be responsible for indications of high precession, such as those seen for GW190521. Thus, our analysis reveals no clear correlation between eccentricity and precession.

We find that the main requirement for reproducing evidence of high in-plane spins is noise, but some degree of precession or eccentricity is also required. Our study highlights the challenges of making conclusive statements about the properties of the binary from which a short signal originates, and we hope that future 3G detectors can help to explore this kind of event further. In order to confidently detect precession for an equal mass, high total mass, and highly precessing binary, we require an SNR above 45, which is most likely to occur with 3G detectors. A network of two or more detectors can improve the recovery of all parameters, especially in the case of high SNR detections.

In conclusion, this study provides a semi-systematic analysis of the possible bias and degeneracy of a short signal from an IMBH binary. We have highlighted the need for caution when making conclusions about the properties of a binary from a short signal, and the importance of exploring other parameters to better understand the PE from similar IMBH binaries. However, due to the cost of computational power and time, we cannot do a full systematic analysis for all parameter space. We hope with the future machine learning PE tools, we can do more exhaustive exploration.

\begin{acknowledgments}
We would like to thank Philippe Jetzer, Shubhanshu Tiwari, Maria Haney, Charlie Hoy and Rhys Green for helpful discussions. 

Y. Xu was supported by China Scholarship Council.
E. Hamilton was supported by grant IZCOZ0\_189876 from the Swiss National Science Foundation (SNSF).
The authors are grateful for computational resources provided by Cardiff University and supported by STFC grant ST/I006285/1. 

This research has made use of data, software and/or web tools obtained from the Gravitational Wave Open Science Center (https://www.gw-openscience.org/ ), a service of LIGO Laboratory, the LIGO Scientific Collaboration and the Virgo Collaboration. LIGO Laboratory and Advanced LIGO are funded by the United States National Science Foundation (NSF) as well as the Science and Technology Facilities Council (STFC) of the United Kingdom, the Max-Planck-Society (MPS), and the State of Niedersachsen/Germany for support of the construction of Advanced LIGO and construction and operation of the GEO600 detector. Additional support for Advanced LIGO was provided by the Australian Research Council. Virgo is funded, through the European Gravitational Observatory (EGO), by the French Centre National de Recherche Scientifique (CNRS), the Italian Istituto Nazionale di Fisica Nucleare (INFN) and the Dutch Nikhef, with contributions by institutions from Belgium, Germany, Greece, Hungary, Ireland, Japan, Monaco, Poland, Portugal, Spain.

All the code to perform the analysis in this paper can be found on \cite{gwutil}

\end{acknowledgments}

\appendix

\section{Eccentricity estimator}\label{sec:eccentricity estimator}

In general relativity, eccentricity cannot be uniquely defined \cite{Ramos-Buades:2018azo}. 
We must therefore choose a consistent method of estimating the degree of eccentricity present in these NR waveforms.
There are a number of different methods to do this \cite{Buonanno:2006ui, Husa:2007rh, Mroue:2010re, Ramos-Buades:2019uvh}. 
Here we use the estimator defined in \cite{Ramos-Buades:2019uvh} because it does not rely on a quasi-circular fit of the orbital frequency.
The eccentricity at time $t$ is defined as
\begin{equation}
	e_{\omega}(t) = \frac{\sqrt{\omega_p^{NR}} - \sqrt{\omega_a^{NR}}}{\sqrt{\omega_p^{NR}} + \sqrt{\omega_a^{NR}}},
\end{equation}
where $\omega_{a,p}^{NR}$ are the GW frequency at apastron and periastron respectively. 
$\omega_{a,p}^{NR}$ are estimated by fitting an ansatz of the form
\begin{equation}
	\omega_{a,p}^{NR} = c_{a,p} \frac{1 + n_{a,p} t}{1 + d_{a,p} t},
\end{equation}
through the value of the GW frequency at the apastron or periastron respectively.
The quantities $c_{a,p}$, $n_{a,p}$ and $d_{a,p}$ are the fit parameters.

\section{NR data injection results}

Here we present the properties of the injections performed in Sec.~\ref{sec:ecc:injection}.
We also give the median and 90\% confidence interval of the results of the injections.
Tab.~\ref{tb:sxs} is for the SXS injections while Tab.~\ref{tb:rit} is for the RIT injections.

\begin{table*}[tb]
\caption{\label{tb:sxs}SXS eccentric injections} 
\begin{ruledtabular}
\begin{tabular}{l|l|l|l|l|l|l|l|l|l|l|l} 
  & $q_\mathrm{ref}$ & $\chi_{\mathrm{eff}_\mathrm{ref}}$ & $\chi_{p_\mathrm{ref}}$ & $e_{t2500}$ & $q$ & $M_\mathrm{tot}$ & $M_\mathrm{chirp}$ & $\chi_\mathrm{eff}$ & $\chi_\mathrm{p}$ & $\rho_\mathrm{p}$ & $\theta_{jn}$\\ [6pt] 
  \colrule
$BBH_{1355}$ & 1.00 & -0.00 & 0.00 & 0.06 & $0.85^{+0.13}_{-0.25}$ & $273.61^{+20.67}_{-19.58}$ & $118.35^{+9.10}_{-10.68}$ & $0.03^{+0.24}_{-0.25}$ & $0.44^{+0.37}_{-0.30}$ & $0.24^{+0.57}_{-0.20}$ & $0.53^{+2.19}_{-0.41}$\\ [6pt] 
$BBH_{1356}$ & 1.00 & 0.00 & 0.00 & 0.08 & $0.86^{+0.12}_{-0.30}$ & $270.17^{+20.46}_{-17.83}$ & $116.74^{+9.40}_{-9.88}$ & $0.03^{+0.24}_{-0.29}$ & $0.39^{+0.45}_{-0.30}$ & $0.24^{+0.54}_{-0.21}$ & $0.57^{+2.25}_{-0.45}$\\ [6pt] 
$BBH_{1357}$ & 1.00 & -0.00 & 0.00 & 0.10 & $0.84^{+0.14}_{-0.23}$ & $269.00^{+20.01}_{-17.77}$ & $116.18^{+9.10}_{-9.22}$ & $-0.01^{+0.22}_{-0.26}$ & $0.42^{+0.36}_{-0.30}$ & $0.24^{+0.56}_{-0.20}$ & $0.53^{+2.30}_{-0.40}$\\ [6pt] 
$BBH_{1358}$ & 1.00 & -0.00 & 0.00 & 0.10 & $0.87^{+0.12}_{-0.23}$ & $272.75^{+23.40}_{-17.97}$ & $117.97^{+10.40}_{-8.27}$ & $0.02^{+0.26}_{-0.23}$ & $0.38^{+0.42}_{-0.27}$ & $0.23^{+0.55}_{-0.20}$ & $0.59^{+2.23}_{-0.44}$\\ [6pt] 
$BBH_{1359}$ & 1.00 & -0.00 & 0.00 & 0.11 & $0.85^{+0.13}_{-0.25}$ & $269.71^{+17.28}_{-15.76}$ & $116.64^{+7.75}_{-8.75}$ & $0.00^{+0.21}_{-0.21}$ & $0.37^{+0.42}_{-0.27}$ & $0.22^{+0.48}_{-0.18}$ & $0.56^{+2.30}_{-0.42}$\\ [6pt] 
$BBH_{1360}$ & 1.00 & -0.00 & 0.00 & 0.15 & $0.83^{+0.15}_{-0.21}$ & $271.16^{+18.60}_{-17.23}$ & $117.18^{+8.26}_{-8.60}$ & $0.04^{+0.21}_{-0.27}$ & $0.42^{+0.37}_{-0.30}$ & $0.26^{+0.57}_{-0.21}$ & $0.61^{+2.23}_{-0.44}$\\ [6pt] 
$BBH_{1361}$ & 1.00 & -0.00 & 0.00 & 0.16 & $0.88^{+0.11}_{-0.24}$ & $273.05^{+19.98}_{-16.62}$ & $118.37^{+8.14}_{-8.22}$ & $0.03^{+0.20}_{-0.20}$ & $0.34^{+0.44}_{-0.24}$ & $0.20^{+0.47}_{-0.16}$ & $0.62^{+2.27}_{-0.45}$\\ [6pt] 
$BBH_{1362}$ & 1.00 & 0.00 & 0.00 & 0.20 & $0.88^{+0.11}_{-0.23}$ & $266.89^{+20.56}_{-17.81}$ & $115.59^{+9.10}_{-8.36}$ & $-0.05^{+0.26}_{-0.23}$ & $0.39^{+0.40}_{-0.29}$ & $0.25^{+0.60}_{-0.21}$ & $0.67^{+2.20}_{-0.51}$\\ [6pt] 
$BBH_{1363}$ & 1.00 & 0.00 & 0.00 & 0.21 & $0.86^{+0.13}_{-0.27}$ & $270.08^{+18.68}_{-17.90}$ & $116.81^{+8.29}_{-9.63}$ & $0.01^{+0.24}_{-0.27}$ & $0.41^{+0.43}_{-0.31}$ & $0.22^{+0.53}_{-0.18}$ & $0.64^{+2.29}_{-0.51}$\\ [6pt] 
$BBH_{0832}$ & 2.00 & 0.01 & 0.80 & 0.02 & $0.63^{+0.21}_{-0.23}$ & $271.26^{+25.67}_{-23.46}$ & $114.17^{+12.31}_{-16.40}$ & $0.10^{+0.23}_{-0.21}$ & $0.32^{+0.36}_{-0.23}$ & $0.28^{+0.56}_{-0.21}$ & $0.69^{+2.09}_{-0.40}$\\ [6pt] 
$BBH_{1364}$ & 2.00 & 0.00 & 0.00 & 0.05 & $0.54^{+0.24}_{-0.18}$ & $272.87^{+20.78}_{-20.62}$ & $112.02^{+12.24}_{-14.20}$ & $-0.03^{+0.22}_{-0.29}$ & $0.32^{+0.41}_{-0.23}$ & $0.43^{+0.93}_{-0.32}$ & $0.94^{+1.62}_{-0.44}$\\ [6pt] 
$BBH_{1365}$ & 2.00 & 0.00 & 0.00 & 0.06 & $0.53^{+0.25}_{-0.21}$ & $269.98^{+23.55}_{-24.39}$ & $110.61^{+13.70}_{-18.08}$ & $-0.02^{+0.21}_{-0.29}$ & $0.34^{+0.37}_{-0.24}$ & $0.43^{+0.75}_{-0.31}$ & $0.90^{+1.76}_{-0.44}$\\ [6pt] 
$BBH_{1366}$ & 2.00 & 0.00 & 0.00 & 0.10 & $0.57^{+0.23}_{-0.20}$ & $286.92^{+34.22}_{-28.49}$ & $119.02^{+17.36}_{-18.17}$ & $0.13^{+0.28}_{-0.26}$ & $0.32^{+0.45}_{-0.23}$ & $0.32^{+0.63}_{-0.24}$ & $0.88^{+1.78}_{-0.47}$\\ [6pt] 
$BBH_{1367}$ & 2.00 & -0.00 & 0.00 & 0.10 & $0.56^{+0.22}_{-0.19}$ & $281.37^{+32.06}_{-25.60}$ & $116.24^{+17.18}_{-17.69}$ & $0.04^{+0.25}_{-0.24}$ & $0.35^{+0.49}_{-0.28}$ & $0.38^{+0.81}_{-0.31}$ & $0.87^{+1.74}_{-0.38}$\\ [6pt] 
$BBH_{1368}$ & 2.00 & -0.00 & 0.00 & 0.10 & $0.55^{+0.20}_{-0.16}$ & $268.31^{+20.35}_{-18.58}$ & $110.73^{+10.85}_{-12.79}$ & $-0.05^{+0.19}_{-0.25}$ & $0.29^{+0.39}_{-0.22}$ & $0.39^{+0.79}_{-0.30}$ & $0.90^{+1.64}_{-0.41}$\\ [6pt] 
$BBH_{1369}$ & 2.00 & 0.00 & 0.00 & 0.20 & $0.59^{+0.23}_{-0.20}$ & $279.26^{+27.85}_{-25.76}$ & $116.75^{+13.25}_{-16.94}$ & $0.06^{+0.24}_{-0.23}$ & $0.30^{+0.42}_{-0.22}$ & $0.30^{+0.70}_{-0.23}$ & $0.77^{+1.82}_{-0.37}$\\ [6pt] 
$BBH_{1370}$ & 2.00 & 0.00 & 0.00 & 0.18 & $0.54^{+0.22}_{-0.19}$ & $273.75^{+29.60}_{-24.22}$ & $112.21^{+15.41}_{-17.10}$ & $0.03^{+0.24}_{-0.26}$ & $0.26^{+0.44}_{-0.20}$ & $0.31^{+0.77}_{-0.25}$ & $0.83^{+1.76}_{-0.41}$\\ [6pt] 
$BBH_{1371}$ & 3.00 & 0.00 & 0.00 & 0.06 & $0.41^{+0.23}_{-0.17}$ & $274.96^{+28.10}_{-28.14}$ & $106.88^{+18.36}_{-22.79}$ & $-0.05^{+0.21}_{-0.29}$ & $0.26^{+0.41}_{-0.19}$ & $0.39^{+0.98}_{-0.30}$ & $1.16^{+1.36}_{-0.59}$\\ [6pt] 
$BBH_{1372}$ & 3.00 & 0.00 & 0.00 & 0.09 & $0.49^{+0.25}_{-0.19}$ & $299.51^{+35.96}_{-35.12}$ & $120.89^{+19.60}_{-24.81}$ & $0.04^{+0.27}_{-0.27}$ & $0.32^{+0.50}_{-0.23}$ & $0.33^{+0.80}_{-0.25}$ & $2.02^{+0.59}_{-1.47}$\\ [6pt] 
$BBH_{1373}$ & 3.00 & 0.00 & 0.00 & 0.09 & $0.43^{+0.23}_{-0.17}$ & $277.26^{+30.60}_{-29.48}$ & $108.48^{+18.42}_{-23.11}$ & $-0.02^{+0.23}_{-0.34}$ & $0.28^{+0.45}_{-0.21}$ & $0.41^{+1.10}_{-0.31}$ & $1.07^{+1.42}_{-0.51}$\\ [6pt] 
$BBH_{1374}$ & 3.00 & -0.00 & 0.00 & 0.18 & $0.49^{+0.22}_{-0.21}$ & $305.11^{+41.77}_{-40.67}$ & $123.04^{+20.89}_{-28.62}$ & $0.10^{+0.36}_{-0.28}$ & $0.37^{+0.48}_{-0.27}$ & $0.36^{+0.79}_{-0.27}$ & $1.08^{+1.51}_{-0.53}$ 
 \\ \end{tabular}
\end{ruledtabular}
\end{table*}

\begin{table*}[tb]
\caption{\label{tb:rit}RIT eccentric injections} 
\begin{ruledtabular}
\begin{tabular}{l|l|l|l|l|l|l|l|l|l} 
  & $q_\mathrm{ref}$ & $\chi_{\mathrm{eff}_\mathrm{ref}}$ & $\chi_{p_\mathrm{ref}}$ & $e_{r}$ & $q$ & $M_\mathrm{tot}$ & $M_\mathrm{chirp}$ & $\chi_\mathrm{eff}$ & $\chi_\mathrm{p}$ \\ [6pt] 
  \colrule
$eBBH_{1318}$ & 1.00 & 0.00 & 0.00 & 0.75 & $0.80^{+0.18}_{-0.31}$ & $275.50^{+28.25}_{-27.09}$ & $118.27^{+12.70}_{-14.77}$ & $-0.10^{+0.29}_{-0.33}$ & $0.47^{+0.39}_{-0.34}$ \\ [6pt] 
$eBBH_{1639}$ & 1.00 & 0.00 & 0.70 & 0.75 & $0.81^{+0.17}_{-0.27}$ & $259.57^{+24.54}_{-20.61}$ & $111.68^{+11.15}_{-10.50}$ & $-0.17^{+0.31}_{-0.33}$ & $0.55^{+0.33}_{-0.38}$ \\ [6pt]
$eBBH_{1318}$ with noise & 1.00 & 0.00 & 0.00 & 0.75 & $0.79^{+0.19}_{-0.26}$ & $280.87^{+30.65}_{-26.39}$ & $120.62^{+13.15}_{-13.11}$ & $-0.00^{+0.37}_{-0.37}$ & $0.59^{+0.31}_{-0.41}$  \\ [6pt] 
$eBBH_{1639}$ with noise & 1.00 & 0.00 & 0.70 & 0.75 & $0.68^{+0.25}_{-0.22}$ & $272.61^{+27.67}_{-24.01}$ & $115.65^{+12.08}_{-12.77}$ & $-0.04^{+0.33}_{-0.31}$ & $0.64^{+0.28}_{-0.40}$
 \end{tabular}
\end{ruledtabular}
\end{table*}

\bibliographystyle{apsrev4-1}
\bibliography{ref}

\end{document}